\documentclass[tradiabstract]{aa}

\usepackage{graphicx}
\usepackage{longtable}
\usepackage{pdflscape}
\usepackage{txfonts}

\newcommand{\au}{{\sc au}}
\newcommand{\micron}{$\mu$m}
\newcommand{\Mjup}{M$_\mathrm{J}$}
\newcommand{\Msun}{M$_\odot$}
\renewcommand{\ion}[2]{#1\,{\sc #2}}
\newcommand{\etal}{et al.}
\newcommand{\Ks}{$\rm K_s$}

\newcommand{\J}{J}
\renewcommand{\H}{H}
\newcommand{\NBdeuxdouze}{$\rm NB_{2.12}$}
\newcommand{\NBdeuxdixsept}{$\rm NB_{2.17}$}
\newcommand{\BrG}{$\rm Br\gamma$}
\newcommand{\FeII}{Fe\,{\sc ii}}
\newcommand{\Kp}{K'}
\newcommand{\Hdeux}{$\rm H_2~(v=1$-$0$)}
\renewcommand{\d}{\mathrm{d}}
\newcommand{\e}{\mathrm{e}}

\titlerunning{Deep imaging of close companions to A--F stars}
\authorrunning{Ehrenreich \etal}

\begin{document}

   \title{Deep infrared imaging of close companions\\to austral A- and F-type stars\thanks{Based on observations made with ESO Telescopes at the Paranal Observatory under programme IDs 076.C-0270, 081.C-0653, and 083.C-0151, and on observations obtained at the Canada-France-Hawai`i Telescope (CFHT) which is operated by the National Research Council of Canada, the Institut National des Sciences de l'Univers of the Centre National de la Recherche Scientifique of France, and the University of Hawai`i.}$^,$\thanks{\scriptsize The full version of this paper with the appendices in available on line at \texttt{http://www-laog.obs.ujf-grenoble.fr/$\sim$dehrenre/articles/afsurvey/}}}

  \author{D.~Ehrenreich\inst{1}, A.-M.~Lagrange\inst{1}, G.~Montagnier\inst{2}, G.~Chauvin\inst{1}, F.~Galland\inst{1}, J.-L.~Beuzit\inst{1} \& J.~Rameau\inst{1}} 

   \offprints{D.~Ehrenreich}

   \institute{Laboratoire d'astrophysique de Grenoble, Universit\'e Joseph Fourier, CNRS (UMR 5571), BP 53, 38041 Grenoble cedex 9, France\\
     \email{david.ehrenreich@obs.ujf-grenoble.fr}
     \and
     European Southern Observatory, Alonso de Cordova 3107, Vitacura Casilla 19001, Santiago 19, Chile
  }

   \date{}

   \abstract{The search for substellar companions around stars with different masses along the main sequence is critical to understand the different processes leading to the formation of low-mass stars, brown dwarfs, and planets. In particular, the existence of a large population of low-mass stars and brown dwarfs physically bound to early-type main-sequence stars could imply that the massive planets recently imaged at wide separations (10--100~\au) around A-type stars are disc-born objects in the low-mass tail of the binary distribution and are thus formed via gravitational instability rather than by core accretion. Our aim is to characterize the environment of early-type main-sequence stars by detecting substellar companions between 10 and 500~\au. The sample stars are also surveyed with radial velocimetry, providing a way to determine the impact of the imaged companions on the presence of planets at $\la10$~\au. High contrast and high angular resolution near-infrared images of a sample of 38 southern A- and F-type stars have been obtained between 2005 and 2009 with the instruments NaCo on the Very Large Telescope and PUEO on the Canada-France-Hawai`i Telescope. Direct and saturated imaging were used in the \J\ to \Ks\ bands to probe the faint circumstellar environments with contrasts of $\sim 5\times10^{-2}$ to $10^{-4}$ at separations of $0\farcs2$ and $1\arcsec$, respectively. Using coronagraphic imaging, we achieved contrasts between $10^{-5}$ and $10^{-6}$ at separations $>5$\arcsec. Multi-epoch observations were performed to discriminate comoving companions from background contaminants. This survey is sensitive to companions of A and F stars in the brown dwarf to low-mass star mass regime. About 41 companion candidates were imaged around 23 stars. Follow-up observations for 83\% of these stars allowed us to identify a large number of background contaminants. We report the detection of 7 low-mass stars with masses between 0.1 and 0.8~\Msun\ in 6 multiple systems: the discovery of a M2 companion around the A5V star \object{HD~14943} and the detection of \object{HD~41742B} around the F4V star \object{HD~41742} in a quadruple system; we resolve the known companion of the F6.5V star \object{HD~49095} as a short-period binary system composed by 2 M/L dwarfs. We also resolve the companions to the astrometric binaries \object{$\iota$~Crt} (F6.5V) and \object{26~Oph} (F3V), and identify a M3/M4 companion to the F4V star \object{$o$~Gru}, associated with a X-ray source. The global multiplicity fraction measured in our sample of A and F stars is $\ge 16\%$. This has a probable impact on the radial velocity measurements performed on the sample stars.}


   \maketitle
%

\section{Introduction}
\label{sec:intro}
 
Stars do not form alone, as they are often found in multiple systems, commonly consisting of two or more stellar components, brown dwarfs, or planets. The various properties of these systems in terms of mass ratios and separations call for different formation mechanisms. 
At the lower end of the mass distribution, more than 400 extrasolar planets have been detected, most of them ($\sim 80\%$) with velocimetric surveys and in most cases the planet-host stars are of solar type. This technique brought observational evidence that giant gaseous planets at separations $\la 6$~\au, which correspond to the largest revolution periods probed by radial velocimetry today, are preferentially forming through rapid accretion of gas on  pre-existing rocky cores, massive enough for triggering the runaway accretion of hydrogen (see, e.g., Mordasini \etal\ 2009). On the other hand, $\sim3\%$ of planets are detected by direct imaging searches at larger separations ($\ga 10$~\au), preferentially around early-type stars. Most of these imaged planets seem to have a different origin, namely the fragmentation of the protoplanetary gaseous circumstellar disc (see, e.g., Dodson-Robinson \etal\ 2009; Kratter, Murray-Clay \& Youdin 2010). Thus, there could be two possible paths of planet formation, depending on the stellar mass and planet separation.

The core-accretion theory (Pollack \etal\ 1996) is currently prefered to the gravitational instability theory (Boss 1997) as the origin mechanism of most planets detected by velocimetry because the latter mechanism is not expected to produce giant planets at distances $\la10$~\au\ (Rafikov 2005). Meanwhile, observational support has been gathered in favour of the core accretion theory: the correlation between the planet frequency and the host star metallicity (Santos, Israelian \& Mayor 2001, 2004) and the emerging fact that planets with an intermediate mass between Neptune's and Saturn's are exceedingly rare are both predicted in this theoretical frame. The recent extent of velocimetric surveys to low-mass main sequence stars (M dwarfs; Mayor \etal\ 2009a,b) yielded new key evidence in favour of this formation mechanism: the fact, for instance, that low-mass planets (about the mass of Neptune) are frequent while gas giants are seldom found around low-mass stars observed by radial velocity or microlensing techniques (Bonfils \etal\ 2007; Sumi \etal\ 2010). On the contrary, the frequency of giant planets should not be impacted by the stellar mass in the frame of gravitational instability, as long as circumstellar discs are massive enough to become unstable (Boss 2006).

In fact, the total mass of the circumstellar disc where planets form, is supposed to scale with the mass of the central star. An increase of the disc total mass can lead to an enhanced growth rate for protoplanetary cores in the disc midplane (Ida \& Lin 2004), leading to a larger number of massive planets around early-type stars than around Sun-like stars. Hence, there should be an observational correlation between planet occurrence and stellar mass (Laughlin, Bodenheimer \& Adams 2004; Ida \& Lin 2005; Kennedy \& Kenyon 2007). Surveying early-type, main-sequence stars hotter and more massive than the Sun should allow to test such a correlation. 

Radial velocimetry can be used to detect planets at short periods around main-sequence early-type stars, as shown by Lagrange \etal\ (2009a). These authors used the HARPS spectrograph installed at the ESO 3.6-m telescope in La Silla (Chile) to survey a sample of 185 stars. They measured on each star the achievable detection limits, taking into account the `jitter'\footnote{The excess of scatter in radial velocity measurements resulting from inhomogeneities on the stellar surface and stellar envelope pulsations.} level of each object. These authors estimated that planets with periods up to 100 days could be found around $\sim50\%$ of the surveyed stars. Constraining the presence of planets at larger separations however requires a different detection technique.

In this respect, direct imaging is a powerful tool used in complement to radial velocimetry for detecting companions at separations typically $\ga 10$~\au. It is sensitive to intrinsically bright objects: low-mass stars, brown dwarfs, and young giant planets, and bring essential insights on the possible origins of these objects (Chauvin \etal\ 2010). Direct near-infrared imaging allowed observers to detect several kinds of substellar companions such as brown dwarfs (e.g., Nakajima \etal\ 1995; Lowrance \etal\ 1999, 2000; Chauvin \etal\ 2005a), objects at the transition between brown dwarfs and planets (e.g., Chauvin \etal\ 2005b), or planetary-mass objects such as \object{2M1207b} in orbit around a brown dwarf (Chauvin \etal\ 2004, 2005c); all of which are likely not formed through core accretion like planets detected by velocimetry, but rather as multiple stars via a cloud fragmentation process or as brown dwarfs through disc fragmentation (see, e.g., Lodato, Delgado-Donate \& Clarke 2005). These detections nevertheless show that star-forming mechanisms can be efficient down to planetary masses.

Recent breakthroughs in high-contrast imaging enriched this picture, as giant planets were directly imaged around  \object{Fomalhaut} (Kalas \etal\ 2008), \object{HR~8799} (Marois \etal\ 2008), and  \object{$\beta$~Pictoris} (Lagrange \etal\ 2009b, 2009c, 2010); all these stars being young and early main-sequence A stars. The discovery of a companion to $\beta$~Pic, a $9\pm3$-\Mjup\ planet with a semi-major axis of $\sim8$ to $15$~\au, implies that giant planets can indeed form in $\sim 10$~Myr. According to theoretical prediction of Kennedy \& Kenyon (2007), this particular planet is close enough to its host star and could have formed in-situ by core accretion. This is not so clear for Fomalhaut~b, which is located 115~\au\ away from its star, meaning that a particular migration scenario is required if the planet formed closer to the star via core accretion (Crida, Masset \& Morbidelli 2009), or for the three planets or brown dwarfs located at 68, 38, and 24~\au\ from HR~8799. Gravitational instability, rather than core accretion, seems more suited to explain the existence of such massive planets on wide orbits (Dodson-Robinson \etal\ 2009). The picture is however not that simple, as the gravitational instabilities leading to the fragmentation of massive circumstellar discs seem to produce objects with masses typically above the deuterium-burning planetary-mass limit. In fact, according to Kratter, Murray-Clay \& Youdin (2010), atypical disc conditions are required to form planetary-mass object via this mechanism. These authors suggest that if these planets formed this way, they must lie in the low-mass tail of the disc-born binary distribution. In this case, a larger number of brown dwarfs or low-mass stars (such as M stars) should be found around A-type stars at distances of 50--150~\au.

In this frame, we perfomed a deep imaging survey of early-type A and F stars included in the velocimetric survey of Lagrange \etal\ (2009a). This study would allow us to test whether low-mass stars and brown dwarfs commonly cohabit with massive main-sequence stars, bringing new constraints on the origin of the massive planets imaged around these stars. In addition, this imaging survey could help determining the impact of stellar multiplicity on the presence of closer-in planets, detected in parallel with radial velocimetry. Binarity is indeed another critical parameter for the theory of formation and evolution of planets. In particular, the separation of the binaries could impact the way giant planets form, either by core accretion or disc instability (Zucker \& Mazeh 2002; Eggenberger, Udry \& Mayor 2004; Desidera \& Barbieri 2007; Duch\^ene \etal\ 2010), and the models of binary discs now predict observational effects. Mayer \etal\ (2005), for instance, predict a lack of planets around each component of binary systems with separations $\la 100$~\au\ because of instabilities in the circumstellar discs. The subsequent dynamical evolution also depends on the properties of the star and the outer companion (see, e.g., Rivera \& Lissauer 2000).
 
Only adaptive optics (AO) deep imaging allows to test properly the presence of massive substellar companions. A number of studies set to probe the existence and the impact of such objects to exoplanetary systems detected by velocimetry has been previously undertook (Patience \etal\ 2002; Luhman \& Jayawardhana 2002; Chauvin \etal\ 2006, 2010; Mugrauer, Seifahrt \& Neuh\"auser 2007; Eggenberger \etal\ 2007). For instance, Eggenberger \etal\ (2007) searched for bright long-period companions around 130 G- and K-type stars and measured a binary fraction of $(8.8\pm3.5)\%$ in a sub-sample of 57 stars with planets detected with radial velocimetry at separations of 0\arcsec8--6\arcsec5 (i.e., 30--250~\au, whereas they found a higher binary fraction of $(12.3\pm3.2)\%$ in a control sub-sample of 73 stars with no planet found. Previous works were also targeted on particular objects, such as \object{Vega} or \object{$\zeta$~Virginis} (Hinkley \etal\ 2010 and references therein), in search of substellar companions.

In this work, we are focusing on the close environment of a sample of austral A- and F-type stars, which we mainly observed from the southern hemisphere using the AO system NaCo installed at the ESO Very Large Telescope (VLT) on Cerro Paranal (Chile). Some austral stars were also observed from the northern hemisphere using PUEO, the AO bonnette of the Canada-France-Hawai`i Telescope (CFHT) on Mauna Kea (USA). In a few cases, we also used NaCo to observe stars with positive declinations. This paper reports on these observations, which are described in Sect.~\ref{sec:obs}. The data reduction is detailed in Sect.~\ref{sec:data}, we present our analysis in Sect.~\ref{sec:results} and discuss the results in Sect.~\ref{sec:discussion}.

\section{Star sample and observations}
\label{sec:obs}

\subsection{Sample selection and biases}

\label{sec:biases}
\begin{figure}
\begin{center}
\resizebox{\columnwidth}{!}{\includegraphics{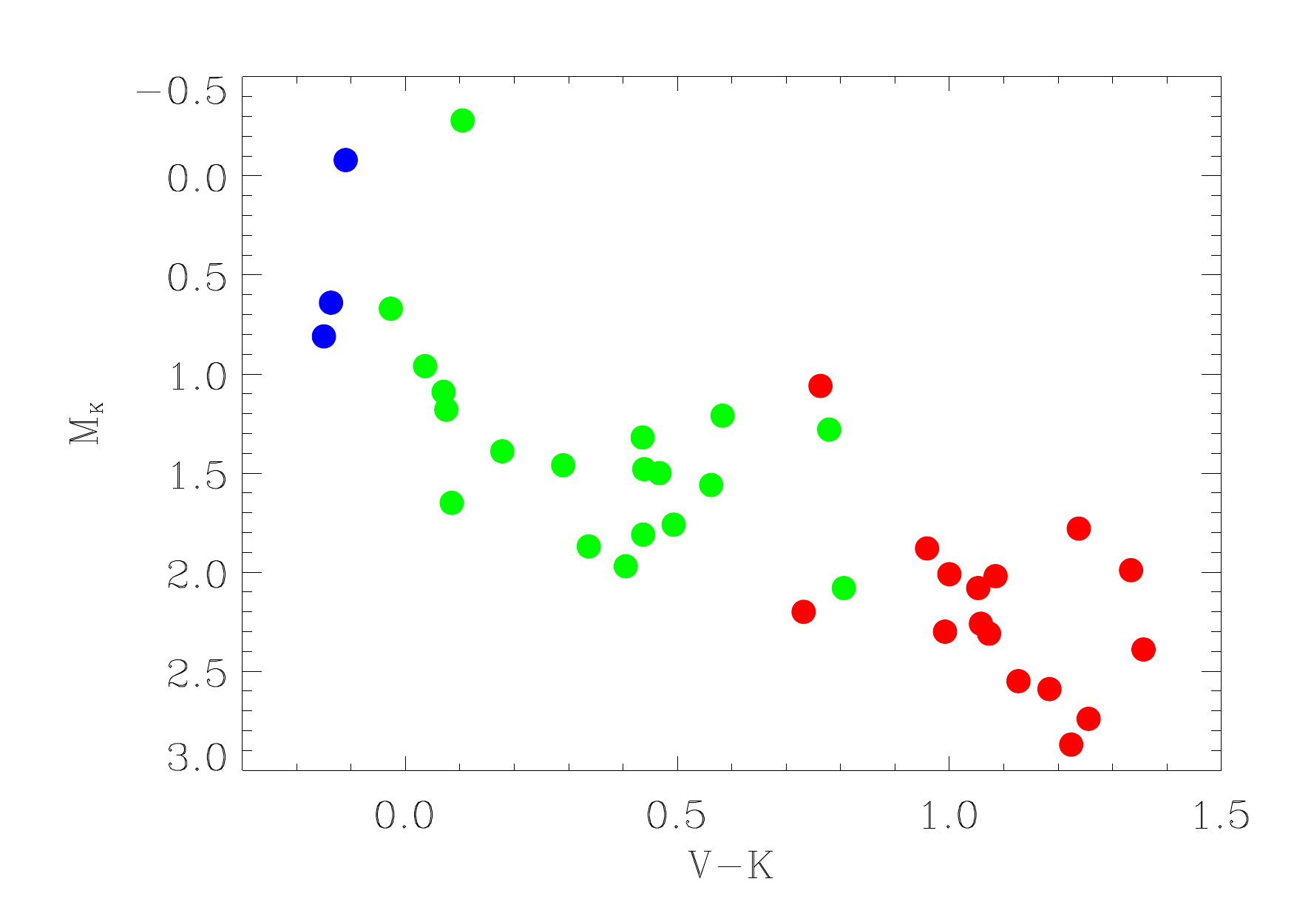}} \caption{\label{fig:colours}Colours of stars in the imaging sample. Stars with spectral types F, A, and B are represented red-, green-, and blue-filled circles, respectively.}
\end{center}
\end{figure}

\begin{figure}
\begin{center}
\resizebox{\columnwidth}{!}{\includegraphics{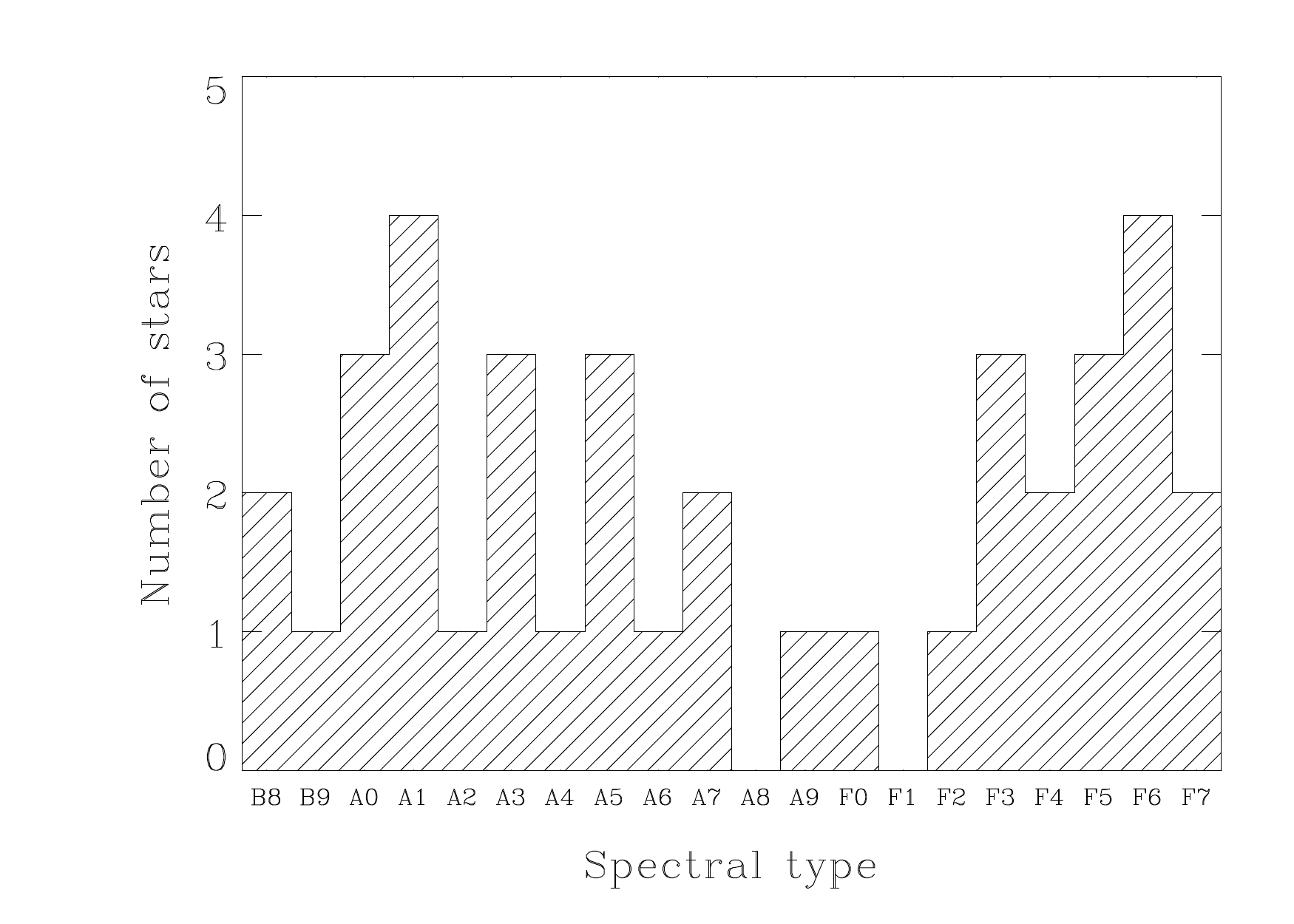}} \caption{\label{fig:sptype}Spectral types of the stars in the sample.}
\end{center}
\end{figure}

The present survey is the imaging part of the radial-velocity survey described in Lagrange \etal\ (2009a) that was designed to detect planets around early-type stars with the HARPS spectrograph at the ESO 3.6-m telescope in La Silla (Chile). The survey is limited to dwarfs with spectral types ranging from F7 to B8\footnote{Velocimetric detection limits for stars earlier than B8 does not fall into the planet domain.}. Figure~\ref{fig:colours} presents the stars in a Hertzsprung-Russell diagram while Fig.~\ref{fig:sptype} shows the number of stars observed per spectral type. In total, we surveyed 38 stars, including 16 F-, 19 A-, and 3 late B-type stars.

\begin{figure}
\begin{center}
\resizebox{\columnwidth}{!}{\includegraphics{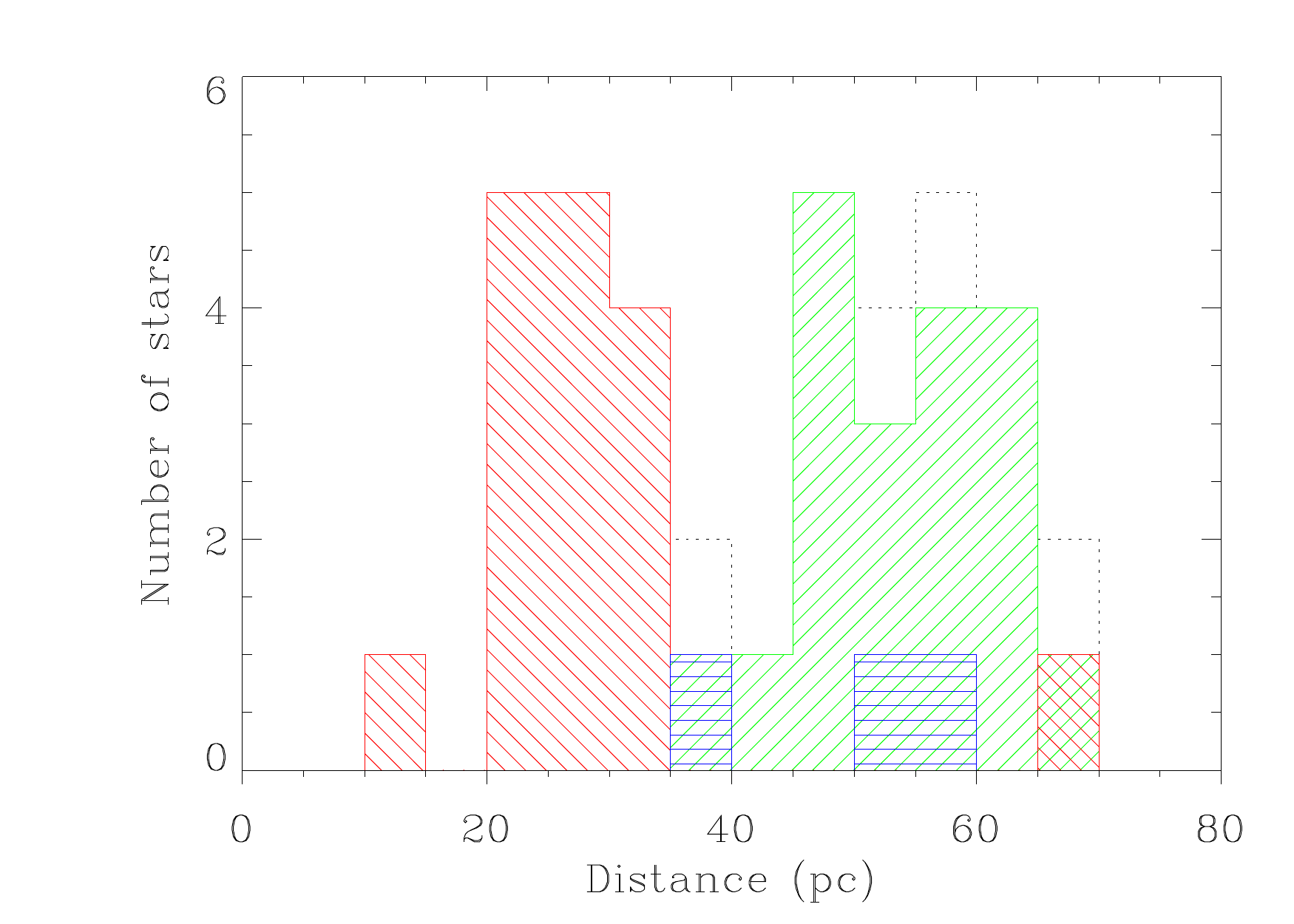}} \caption{\label{fig:distances}Number of stars per 5-pc distance bin. The F, A, and B stars are represented by the histograms filled with tight diagonal red stripes, diagonal green stripes, and horizontal blue stripes, respectively. The dotted line stands for the whole star sample.}
\end{center}
\end{figure}

The velocimetric and imaging surveys are also volume-limited, with distance limits set at 33 and 67~pc for the F0--F7 and B8--A9 dwarfs, respectively. Figure~\ref{fig:distances} gives the number of stars observed per 5-pc distance bin. The difference in distance limits would allow us to have roughly the same number of A and F stars in the sample. Note, however, that one early F star (HD~4293) with a distance of 66.6~pc is included in the survey. 

Finally, the few stars present in our sample that are not included in Lagrange \etal 's (2009a) sample are part of the northern radial-velocity survey in progress with the SOPHIE spectrograph (Bouchy \etal\ 2009) at the 1.93-m telescope in Observatoire de Haute-Provence (France). The properties of all stars in the sample can be found in Table~\ref{tab:sample}. 

\addtocounter{table}{1}

While more details about the sample selection are given in Lagrange \etal\ (2009a), we emphasize three biases. (i) First, spectroscopic binaries and close visual binaries with separations smaller than 5\arcsec\ known at the beginning of the velocimetric survey were excluded from the target list. Hence, these were not considered in the imaging survey as well. (ii) In addition, the present imaging survey is in fact strongly biased towards `interesting' targets, i.e., stars around which companion candidates (CC) were detected during early epochs. This strategy was set mainly in order to compensate for the limited amount of observing time devoted to the imaging programme. (iii) Finally, poor atmospheric conditions and technical problems that were experienced especially during observing runs VI and VII, prevented us from obtaining second-epoch observations for 13\% of those targets with companion candidate(s) identified during a first epoch; atmospheric conditions making the AO correction loop unstable also prevented us from obtaining the highest achievable contrast -- through the use of the coronagraph -- for 10\% of the targets. 


\subsection{Observations}
Data were recorded during seven observing runs (or epochs) performed between January 2005 and August 2009, at the VLT and CFHT, where we totalized $7+8$ nights of observations, respectively. The observing dates, instruments, and filters used during each epoch are summarized in Table~\ref{tab:platescale}, while the detailed observing set-ups for each target are given in Table~\ref{tab:obs}.

\begin{table*}[!t]
\centering
\caption{\label{tab:platescale} Epoch dates and astrometric calibrations for all observing runs.}
\begin{tabular}{ccccccc}
\hline \hline
Epoch     & UT Date              & Nights            & Instrument/Camera & Filter  & Plate scale    & True north     \\
          &                      &                   &                   &         & (mas)          & (\degr)        \\
\hline
I         & 2005-01-27           & 4                 & PUEO/KIR          & \BrG    & $35.06\pm0.1$  & $0\pm0.1$      \\
II        & 2005-11-06           & 3                 & NaCo/S27          & \J      & $27.06\pm0.06$ & $0.02\pm0.1$   \\
          &                      &                   & NaCo/S27          & \Ks     & $27.06\pm0.06$ & $0.02\pm0.1$   \\
III       & 2007-01-28           & 2                 & PUEO/KIR          & \BrG    & $35.08\pm0.03$ & $-2.39\pm0.04$ \\
          &                      &                   & PUEO/KIR          & K       & $34.87\pm0.03$ & $-2.44\pm0.03$ \\
IV        & 2007-11-16           & 2                 & PUEO/KIR          & \BrG    & $35.04\pm0.03$ & $-2.63\pm0.02$ \\
          &                      &                   & PUEO/KIR          & \FeII   & $34.83\pm0.01$ & $-2.64\pm0.02$ \\
V         & 2008-08-20           & 2                 & NaCo/S27          & \Ks     & $27.08\pm0.03$ & $0.04\pm0.09$  \\
          &                      &                   & NaCo/S13          & \Ks     & $13.27\pm0.04$ & $-0.04\pm0.22$ \\
          &                      &                   & NaCo/S13          & \H      & $13.26\pm0.04$ & $-0.02\pm0.21$ \\
VI        & 2009-04-26           & 1                 & NaCo/S27          & \Ks     & $27.04\pm0.02$ & $-0.41\pm0.06$ \\
          &                      &                   & NaCo/S13          & \H      & $13.19\pm0.06$ & $-0.88\pm0.37$ \\
          &                      &                   & NaCo/S13          & \J      & $13.19\pm0.05$ & $-0.74\pm0.26$ \\
VII       & 2009-08-27           & 1                 & NaCo/S27          & \Ks     & $27.06\pm0.02$ & $-0.36\pm0.05$ \\
\hline
\end{tabular}
\end{table*}

\addtocounter{table}{1}

\subsubsection{VLT/NaCo}
\label{sec:obs_naco}
We used the NAOS-CONICA instrument (NaCo) set on the VLT Unit Telescope 4 (Yepun) to benefit from both the high image quality provided by the Nasmyth Adaptive Optics System (NAOS; Rousset \etal\ 2003) at infrared wavelengths and the good dynamics offered by the Near-Infrared Imager and Spectrograph (CONICA; Lenzen \etal\ 2003) detector, in order to study the close circumstellar environment of 38 early-type stars. The NaCo Shack-Hartmann visible wavefront sensor was chosen to perform the AO corrections on these bright targets, used as self-references. NaCo allowed us to perform direct imaging as well as coronagraphic imaging in order to improve the image contrast. We used the coronagraphic mode consisting in a Lyot stop in the pupil plan of the telescope, combined with a occulting mask of diameter $\diameter=0\farcs7$ inserted in the focal plane. According to the atmospheric conditions, we used  different broad- and narrow-band filters whose properties (central wavelength $\lambda_c$, bandpass $\Delta\lambda$, and transmission $T$) are listed in Table~\ref{tab:filters}. For instance, poor atmospheric conditions degrade the Strehl ratio $S$; using a long-wavelength filter (typically \Ks) allows to compensate for this degradation since the value of $S$ also scales with wavelength as $S = \exp -(2\pi\omega/\lambda)^2$, where $\omega$ is the root-mean-square deviation of the wavefront.

Two objectives were employed in order to optimize the point spread function (PSF) sampling in these different bandpasses. The S13 camera has a field of view of $14\times14$~arcsec$^2$ and a mean plate scale of 13.21~mas pixel$^{-1}$; it was preferentially used when observing in \J\ and \H\ bands. When observing in the \Ks\ band, the S27 camera was chosen: it offers a $28\times28$~arcsec$^2$ field of view and a mean plate scale of 27.06~mas~pixel$^{-1}$. For each epoch, precise plate scales were redetermined using astrometric calibrators (see Table~\ref{tab:platescale}).

\begin{figure*}
\begin{center}
\resizebox{0.45\textwidth}{!}{\includegraphics{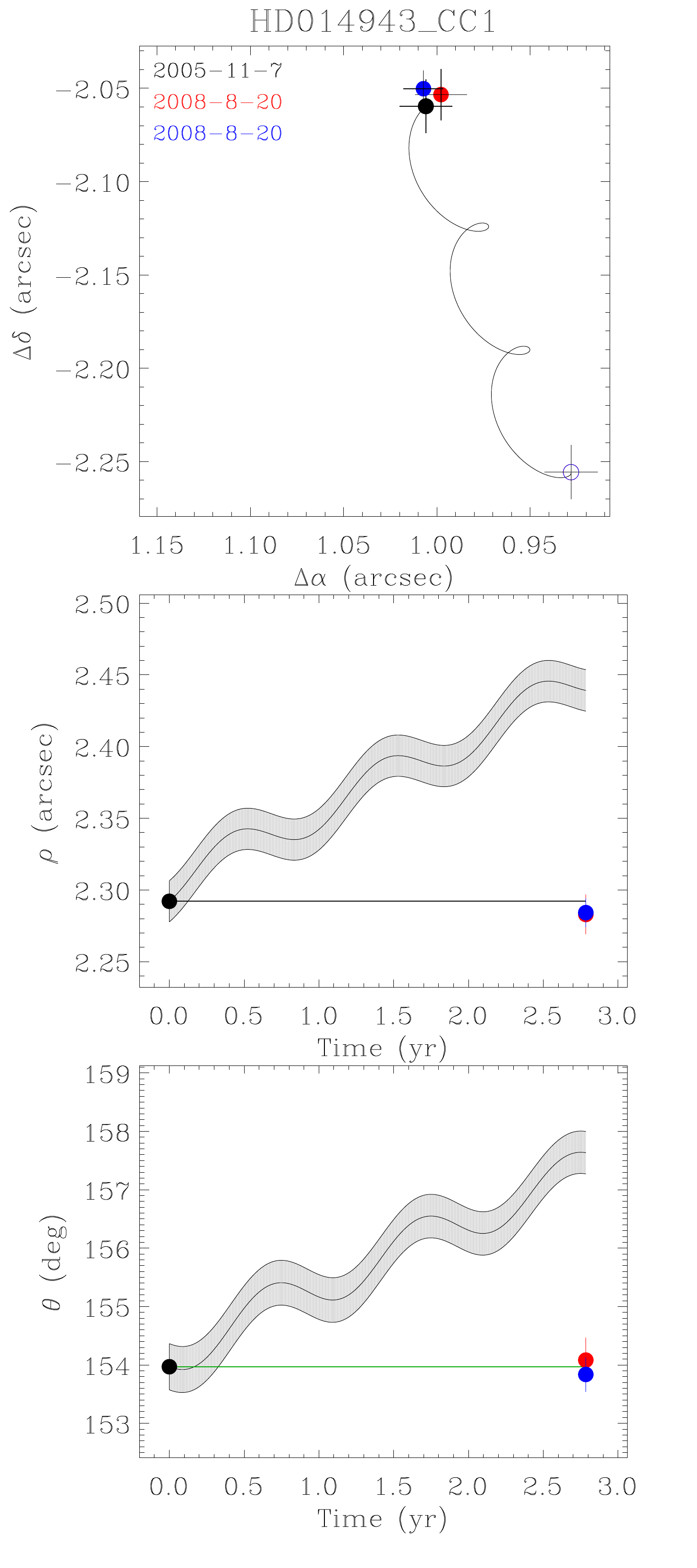}}
\resizebox{0.45\textwidth}{!}{\includegraphics{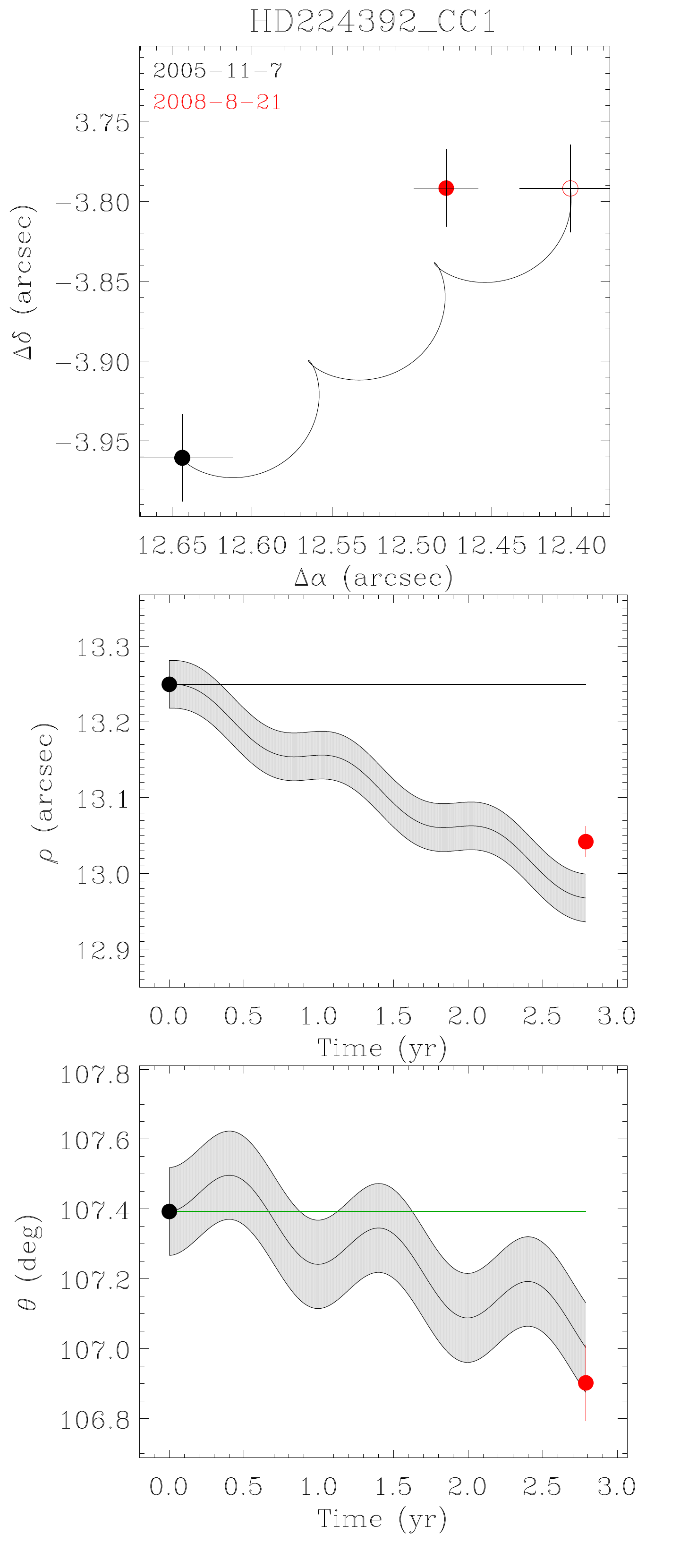}}
\caption{\label{fig:diagastro}Examples of the status determination of companion candidates around HD~14943 (left) and $\eta$~Tuc (HD~224392; right), from astrometric measurements at two different epochs. The positions on the sky are reported in the upper panel: the candidates were first imaged on 2005-11-07 (filled black circles), then on 2008-08-20 (filled red and blue circles; there are two measurements in the case of HD~14943). The positions predicted for the candidates on 2008-08-20 if they were background objects from the field are indicated by the empty red and blue circles, and their trajectories between the two epochs figured by the black curves, which take into account the stellar proper motions and the parallactic motions. The middle and lower panels show the evolutions of the projected separations and the position angles, respectively, with time. One-sigma error bars are represented. A zero or near-zero evolution of the separation and position angle with time is indicative of a physical association. The candidate at the left is bound to its star with indicative probabilities $P_\mathrm{bkg} = 0\%$ and $P_\mathrm{cmv} = 96.7\%$ (following the calculations of Sect.~\ref{sec:companionship}), while the candidate at the right is a background contaminant, with $P_\mathrm{bkg} = 12.0\%$ and $P_\mathrm{cmv} = 0\%$.}
\end{center}
\end{figure*}

Our observing strategy consists, for each target, in obtaining a first-epoch image and, when companion candidates are detected, to perform a second-epoch observation in order to test their status: either field objects (background contaminants) or comoving objects (physically bound to the targeted star). The discrimination between these two possibilities is allowed by the stellar proper and parallactic motions in the plane of the sky. This is illustrated in Fig.~\ref{fig:diagastro}. Hence, depending on motion amplitudes, a few months to a few years are necessary between the first- and second-epoch observations. For instance, for two observations performed three years apart, and a precision on the measured positions of individual point sources on each epoch of $\sim10$--$20$~mas (see Table~\ref{tab:cc}), we can roughly estimate that a 3-$\sigma$ discrimination between comoving and non-comoving sources is possible for typical minimum proper motions of $\sim15$--$30$~mas~yr$^{-1}$.

A target is first observed both in direct imaging and coronagraphic modes. For epochs~V, VI, and VII, the \texttt{CoroObsAstro} observing template was used: this template consists in acquiring first some coronagraphic exposures on the object without any neutral density filter (\texttt{Full\_Uszd} setting, where the pupil area is simply reduced by the presence of the Lyot stop, which blocks 14\% of the light). Then, a neutral density filter (\texttt{ND\_Short}) is inserted in the focal plane while removing both the coronagraphic mask and the Lyot stop in the focal and pupil plan, respectively. A direct image is then taken, the object being at the same exact location as behind the mask. This allows for retrieving the star position behind the mask, enabling precise astrometric measurements of separation with companion candidates only visible in coronagraphic mode. The star is then jittered (by steps of $\sim 5\arcsec$) on 4 different locations on the detector in order to obtain the background sky estimation for the direct images. The last offset sends the telescope $\sim 30\arcsec$ away from the target star, and the occulting mask is used again -- the neutral density filter is removed once more -- as 5 jittered exposures are taken on the sky in order to estimate the sky background for coronagraphic images. This template was not available during our first NaCo observing run (epoch~II), hence the precise location of the star behind the mask is not known, and the astrometry is therefore less precise than for later observations. In all cases, the same amount of time was spent on the object and on the sky. 

A systematic astrometric shift is induced when switching the neutral density filter back and forth: we used pinhole calibrations recorded at each epoch to ensure that this shift is small: typically $\pm0.1$~pixel and $\pm0.01$~pixel along the detector $x$- and $y$-axes, respectively. This shift is included in our error budget within the $\sigma_s$ term of Eq.~(\ref{eq:error}).

When a companion candidate can be detected without the coronagraph, the star is reobserved at a later epoch, only in direct imaging mode. NaCo observing templates \texttt{ImgGenericOffset} and \texttt{ImgAutoJitter} were used for this purpose.

\subsubsection{CFHT/PUEO}
PUEO (Rigaut \etal\ 1998) is the AO bonnette mounted on the 3.6-m CFHT. It was used in combination with the near-infrared camera KIR (Doyon \etal\ 1998) which has a field of view of $35\times35$~arcsec$^2$, and a mean plate scale of $34.8$~mas~pixel$^{-1}$. Depending on atmospheric conditions and source brightnesses, we used broad- and narrow-band filters listed in Table~\ref{tab:filters}.

We performed unsaturated direct imaging to investigate as close as possible to the star, while the detector was saturated in order to increase the contrast at larger separations (there is no occulting mask on PUEO). A special care was taken to ensure that the unsaturated exposures could be used as references to measure the relative position and photometry between the star and the companion candidate(s). Images are jittered on the detector so that the sky background can be retrieved. The observing method is otherwise similar to that described above for NaCo. The 6 stars from our sample with PUEO epochs have all been observed with NaCo, and are thus included in this study for consistency.

\begin{table}[!t]
\begin{center}
\caption{\label{tab:filters} Properties of the filters used during the survey.}
\begin{tabular}{lrrr}
\hline\hline
Filter & $\lambda_c$ ($\mu$m) & $\Delta\lambda$ ($\mu$m) & $T$ (\%) \\
\hline
\multicolumn{4}{c}{VLT/NaCo}                                        \\
\hline
\J             & 1.27         & 0.25                     & 78       \\
\H             & 1.66         & 0.33                     & 77       \\
\NBdeuxdouze   & 2.122        & 0.022                    & 55       \\
\NBdeuxdixsept & 2.166        & 0.023                    & 52       \\
\Ks            & 2.18         & 0.35                     & 70       \\
\hline
\multicolumn{4}{c}{CFHT/PUEO}                                       \\
\hline
\H             & 1.632        & 0.296                    & 85       \\
\FeII          & 1.650        & 0.015                    & 70       \\
\Kp            & 2.115        & 0.350                    & 92       \\
\Hdeux         & 2.117        & 0.020                    & 65       \\
\BrG           & 2.169        & 0.020                    & 70       \\
K              & 2.198        & 0.336                    & 90       \\
\hline
\end{tabular}
\end{center}
\end{table}

\section{Data reduction}
\label{sec:data}

\subsection{Calibrations}
Flat fields have been taken on the twilight sky for each observing run at VLT (usually within a few days from the observing nights). At CFHT, dome flat fields were taken for every filter at least once per run.

A well-known field in the Orion Trapezium centered on the star \object{$\theta^1$~Orionis~C} was used for astrometric calibrations (McCaughrean \& Stauffer 1994). One observation per filter/camera set-up used was taken at each observing epoch to determine the mean plate scale and the true north orientation. On each observation, the exposure time was adjusted to saturate the 2 or 3 brightest stars of the field so that from 15 to 40 stars (depending on the size of the field, from $13\arcsec$ with the S13 camera of NaCo to $35\arcsec$ with PUEO) have a sufficient signal-to-noise ratio ($\geq 5$). The position of the centroid $(s_x,s_y)$ of each non-saturated star on the reduced image of the field is measured in pixels; these values are then compared to the position on the sky $(\rho_x, \rho_y)$ in arcseconds reported by McCaughrean \& Stauffer (1994) with the following equations:
\begin{eqnarray}
\rho_x &=& x_0 + p_x s_x \cos \theta_n  + p_y s_y \sin \theta_n \\
\rho_y &=& y_0 - p_x s_x \sin \theta_n  + p_y s_y \sin \theta_n 
\end{eqnarray}
where $p_x$ ($p_y$) is the plate scale along the $x$- ($y$-) axis, $\theta_n$ is the orientation of the detector on sky, and $x_0$ and $y_0$ are offsets giving the correspondance between the absolute positions (it is only used to solve the equation). A Levenberg-Marquardt minimization is then applied to find the plate scale solution. The errors on the plate scale and orientation are given by the standard deviation of the corresponding parameter. The mean plate scales and orientations are reported in Table~\ref{tab:platescale}.

\subsection{Coronagraphic data}
The reduction procedure makes use of the \texttt{eclipse} utility library (ESO 1996--2002; Devillard 1997). For each observing template on a target, coronagraphic exposures are sorted by eye. Good exposures (i.e., exposures where the star is well occulted and the AO correction loop is closed) are then concatenated into a data cube. A coronagraphic sky is created as the median of all dithered coronagraphic exposures on the sky, and is corrected twice for bad pixels detected (i) on the reduced flat field and (ii) on the median sky once the first bad-pixel rejection as been applied. The cleaned median sky is then subtracted to each plane of the data cube, which are next divided by the flat field. The 2 bad-pixel maps are again used on the flat-fielded data cube. The final reduced image is extracted as the median of the data cube.

\subsection{Direct imaging data}
The idea is similar to the reduction of data taken in coronagraphic mode, except that all the steps are performed by the \texttt{jitter} routine, which realigns dithered direct images via a cross-correlation algorithm and does all the cosmetic tasks. An additional (but similar) reduction is devoted to the first direct image taken after the coronagraph is removed (see Sect.~\ref{sec:obs_naco}); this reduced image will be used as an astrometric reference for the position of the star behind the mask.

\section{Analysis}
\label{sec:results}

\subsection{Extraction of astrometric and photometric parameters of candidate companions}
\label{sec:results:extraction}
Candidate companions are spotted by a close eye-examination of each star, and their absolute position on the detector determined to a $\sim 1$-pixel precision. The position of the star is similarly estimated in unsaturated direct images; it is estimated from the astrometric direct images for coronagraphic observations. These positions are then used as inputs by a dedicated \texttt{IDL} wrapper programme, that successively performs three astrometric and photometric estimations. An aperture estimation (i), where the aperture is centered on the input position, returns the centroid position as the barycentre of the light within the aperture, and an arbitrary magnitude estimation from the integration of all pixels within the aperture, weighted by the flux of the reference PSF within the same aperture. The position $(\Delta x, \Delta y)$ and magnitude $\Delta m$ of each companion candidate are then given relatively to those of the star. A Gaussian fitting (ii) using the \texttt{IDL} routine \texttt{GCNTRD} is also applied to each input position and returns relative separations. Finally, (iii) a deconvolution algorithm using a Levenberg-Marquardt minimization of the log-likelyhood function (V\'eran 1997) is employed: it basically consists in a PSF fitting within the pupil plane. Procedures (i) and (ii) work well when the objects are well separated on the image, and give similar results than procedure (iii). However, they do not give accurate results for tight systems (typically tighter than the aperture radius used). In these cases, our estimations rely on procedure (iii).

The astrometric precision is ultimately limited by the astrometric calibrator, and the precision $\sigma_p$ (in mas) obtained on the estimation of the plate scale $p$ (in mas~pixel$^{-1}$) at each epoch (which are listed in Table~\ref{tab:platescale}). Yet, our procedure for extracting the astrometric parameters of an image is also characterized by an uncertainty $\sigma_s$ on the measured separation $s$ (both in pixels). The total uncertainty on the measured separation $\rho$ (in mas) is thus
\begin{equation} \label{eq:error}
\frac{\sigma_\rho}{\rho} = \sqrt{\left(\frac{\sigma_s}{s}\right)^2 + \left(\frac{\sigma_p}{p}\right)^2}. 
\end{equation}
The values of $\sigma_s$ depend on various parameters such as the actual separation of the components as well as their contrast. In the following, we choose to take $\sigma_s = 0.5$~pixel as a conservative value. In fact, this choice does not impact our conclusions concerning moderately separated systems. Meanwhile, this value can be thought of as representative since it is clear that the real precision we can achieve does not allow us to resolve the most difficult cases, such as the very tight companion to HD~43940.

Since the stellar flux is always estimated from the direct (unsaturated) image, both the direct and coronagraphic images are normalized by their respective exposure times when the photometry is performed for objects present in NaCo coronagraphic images. The magnitude difference is then increased by 4.6~mag to take the transmission difference introduced by the use of a neutral density filter into account. In the case of NaCo coronagraphic imaging, the use of a slightly undersized aperture due to the presence of the Lyot stop in the pupil plan leads to an additional uncertainty of $\sim4\%$ on the fluxes measured when the occulting mask is on, with respect to the full aperture available when it is off (Boccaletti \etal\ 2007).

As for the separation, the determination of the contrast between two components is also impacted by their respective brightnesses and separation. We can actually estimate our photometric precision from the set of dispersion of the measurements taken on the same objects at different epochs: the median dispersion obtained is 0.4~mag (45\% on the measured fluxes). Astrometric and photometric measurements for all companion candidates imaged are given in Table~\ref{tab:cc}. Stars without any detected companion candidates are listed in Table~\ref{tab:nodet}.

\addtocounter{table}{1}

\begin{table*}
\begin{center}
\caption{\label{tab:nodet} Sample stars around which no companion candidates were detected.\newline
*At CFHT, saturated imaging was used instead of coronagraphic imaging.}
\begin{tabular}{rl*{7}{c}}
\hline\hline
\#& Star name   & Date       & Filter       & Camera & \multicolumn{2}{c}{$\rm DIT~(s) \times NDIT \times NEXP$} \\
  &             &            &              &        & Direct imaging        & Coronagraphic imaging \\
\hline
1 & HD   2834   & 2005-11-06 & \Ks          & S27    & $0.8\times5\times10$  & $3.0\times9\times22$  \\
2 & HD   3003   & 2005-11-07 & \Ks          & S27    & $0.8\times5\times10$  & $6.0\times3\times30$  \\
3 & HD   4293   & 2005-11-08 & \Ks          & S27    & $1.0\times5\times10$  & $10.0\times3\times20$ \\
4 & HD  10939   & 2005-11-06 & \Ks          & S27    & $0.4\times5\times10$  & $5.0\times6\times20$  \\
7 & HD  16754   & 2005-11-07 & \Ks          & S27    & $0.4\times5\times10$  & $5.0\times6\times20$  \\
8 & HD  19545   & 2008-08-20 & \Ks          & S27    & $2.0\times5\times10$  &                       \\
12& HD  29992   & 2008-08-20 & \Ks          & S27    & $0.8\times13\times10$ & $2.0\times15\times10$ \\
13& HD  31746   & 2005-11-07 & \Ks          & S27    & $0.6\times5\times10$  & $8.0\times5\times15$  \\
23& HD 112934   & 2007-01-28 & K            & KIR*   & $1.0\times1\times18$  & $30.0\times1\times30$ \\
  &             & 2008-08-20 & \Ks          & S27    & $3.0\times4\times10$  &                       \\
24& HD 116568   & 2008-08-20 & \Ks          & S27    & $1.0\times10\times10$ &                       \\
28& HD 186543   & 2008-08-20 & \Ks          & S27    & $1.0\times12\times10$ & $5.0\times6\times10$  \\
30& HD 200761   & 2008-08-21 & \Ks          & S27    & $1.0\times10\times10$ &                       \\
  &             & 2009-08-27 & \Ks          & S27    & $0.35\times30\times10$& $1.0\times30\times10$ \\
31& HD 209819   & 2008-08-21 & \Ks          & S27    & $0.35\times30\times10$&                       \\
  &             & 2009-08-27 & \Ks          & S27    & $0.35\times30\times10$& $1.0\times30\times10$ \\
34& HD 216627   & 2008-08-21 & \Ks          & S27    & $0.35\times25\times10$&                       \\
  &             & 2009-08-27 & \NBdeuxdouze & S27    & $2.0\times30\times10$ & $1.0\times5\times10$  \\
37& HD 223011   & 2008-08-20 & \Ks          & S27    & $5.0\times2\times10$  &                       \\    
\hline
\end{tabular}
\end{center}
\end{table*}

\subsection{Companionship}
\label{sec:companionship}
The companionship, i.e., the fact that a candidate companion is comoving with the primary star in the vincinity of which it has been imaged -- which is indicative of a physical association bound by gravity -- is appreciated by measuring the evolution of the separation $\rho = \sqrt(\Delta x^2 + \Delta y^2)$ and the position angle $\theta = \arctan \Delta x / \Delta y$ (from north to east) between the candidate and the star over several epochs of observation. The values of $\Delta x$ and $\Delta y$ are corrected for the orientation $\theta_n$ of the detector north with respect to the true north (given in Table~\ref{tab:platescale} at each epoch). The shift in the position (right ascension $\alpha$ and declination $\delta$) of a companion candidate between two epochs $i$ and $j$ is
\begin{eqnarray}
\Delta \alpha_{i \to j} & = & \rho_j \sin \theta_j - \rho_i \sin \theta_i \\
\Delta \delta_{i \to j} & = & \rho_j \cos \theta_j - \rho_i \cos \theta_i.
\end{eqnarray} 
These measured shifts are compared to the theoretical apparent motion the object should have on the sky, given the stellar proper motions listed in Table~\ref{tab:sample}, the parallactic motion, and assuming, as standardly done, that the candidate is a motionless background contaminent. Figure~\ref{fig:diagastro} shows two examples of companionship determination. The positions of all companion candidates obtained during several observing runs are represented in Fig.~A1 (see the Appendices\footnote{\scriptsize The full version of this paper with the appendices in available on line at \texttt{http://www-laog.obs.ujf-grenoble.fr/$\sim$dehrenre/articles/afsurvey/}}).

For observations performed at $N$ epochs, it is possible to calculate the probability $P_\mathrm{bkg}$ that a companion candidate is a background object, using a $\chi^2$ probability test of $2N-2$ degrees of freedom:
\begin{equation} \label{eq:chi2bkg}
  \chi^2_\mathrm{bkg} = \sum_{i=1}^{N} \left[ \frac{\left(\Delta\alpha_{1 \to i} - \Delta\alpha^\star_{1 \to i}\right)^2}{\sigma_{\Delta\alpha_{1 \to i}}^2 + \sigma_{\Delta\alpha^\star_{1 \to i}}^2} +
\frac{\left(\Delta\delta_{1 \to i} - \Delta\delta^\star_{1 \to i}\right)^2}{\sigma_{\Delta\delta_{1 \to i}}^2 + \sigma_{\Delta\delta^\star_{1 \to i}}^2} \right],
\end{equation}
where $\Delta\alpha^\star_{1\to i}$ and $\Delta\delta^\star_{1\to i}$ are the theoretical shifts in $\alpha$ and $\delta$ in the position of the star between epochs 1 and $i$, taking the stellar proper motion and the parallactic motion into account.

The probability of observing a value of $\chi^2$ that is larger than that obtained with Eq.~(\ref{eq:chi2bkg}) for a random sample of $N$ observations with $\nu=2N-2$ degrees of freedom is the integral of the probability density of a $\chi^2$-distribution (Bevington \& Robinson 2003),  
\begin{equation} \label{eq:Pbkg}
P_\chi(\chi^2; \nu) = \frac{1}{2^{\nu/2}\Gamma\left(\nu/2\right)} \int_{\chi^2}^{+\infty}\left(x^2\right)^{(\nu-2)/2}\e^{-(x^2)/2} \d\left(x^2\right)
\end{equation}
where the $\Gamma(n)$ function is equivalent to the factorial function $n!$ extended to nonintegral arguments, and $(x^2)$ represents the possible values of $\chi^2$ within the integral sum. The probability $P_\mathrm{bkg}$ is obtained from Eq.~(\ref{eq:Pbkg}); practically, we are using the IDL \texttt{chisqr\_pdf} function. A status of background contaminant is assigned to each object for which $P_\mathrm{bkg} > 0.01\%$. This probability is given for each object in Fig.~A1. This first test allows to reject 13 out of 41 detected point sources (32\%), identified as background contaminants (labelled `(B)ackground' in Table~\ref{tab:cc}).

Because of systematic effects in the position measurements, due for instance to variations in the stellar proper motions, discrepancies between the real values and those from the literature, or the fact that background objects $\emph{have}$ a non-negligible proper motion on the sky, we calculated $P_\mathrm{bkg} < 0.01\%$ for some objects that are evidently not comoving with the primary star. This is, for instance, the case of HD~68456 CC\#1 or HD~91889 CC\#1 (see Fig.~A1). For this reason, we additionally define the probability $P_\mathrm{cmv}$, also given in Fig.~A1, that a companion candidate is comoving with the star, using a test similar to Eq.~(\ref{eq:chi2bkg}), 
\begin{equation} \label{eq:chi2cmv}
  \chi^2_\mathrm{cmv} = \sum_{i=1}^{N} \left[ \left(\frac{\Delta\alpha_{1 \to i}}{\sigma_{\Delta\alpha_{1 \to i}}}\right)^2 +
\left(\frac{\Delta\delta_{1 \to i}}{\sigma_{\Delta\delta_{1 \to i}}}\right)^2 \right].
\end{equation}
The result is injected into Eq.~(\ref{eq:Pbkg}). Note, however, that the value of $P_\mathrm{cmv}$ \emph{can only be used to give a hint of a physical association}. In fact, this probability does not take into account the orbital motion of a true comoving companion. Hence, an object with very small values of $P_\mathrm{bkg}$ and $P_\mathrm{cmv}$ ($< 0.01\%$) is not necessarily a background contaminant because of orbital motions: this is typically the case of HD~49095 CC\#1 and CC\#2, which are revolving around each others while clearly following the star over the 4 observing epochs. In the following, we consider that an object with $P_\mathrm{bkg} < 0.01\%$ and $P_\mathrm{cmv} > 0.01\%$ is comoving with the star: 4 companion candidates out of 41 detected point sources (10\%) are confirmed comoving objects according to this test (they are labelled `(C)omoving' in Table~\ref{tab:cc}). Consequently, an object with $P_\mathrm{bkg} < 0.01\%$ and $P_\mathrm{cmv} < 0.01\%$ could be considered to be a background contaminant. On this basis, we can confidently reject 6 additional point sources (15\%). 

Another tricky case is that of HD~41742 CC\#1 ($P_\mathrm{bkg}=0\%$, $P_\mathrm{cmv}=0\%$), observed during 4 epochs: as can be seen in Fig.~A1 the second-epoch measurement is far from that expected of a comoving object; it is also far from that of a background contaminant. Meanwhile, measurements recorded at all other epochs are well compatible with a comoving object (see HD~41742 CC\#2 for comparison). This object is considered to be comoving because the measurement made during the second epoch for all three companion candidates to this star seem to be outliers. If we reject the measurements obtained at this epoch for this star, then CC\#1 is comoving while CC\#2 and CC\#3 indeed are background contaminants.

Finally, two point sources observed at several epochs remain with an ambiguous nature (labelled `(A)mbiguous' in Table~\ref{tab:cc}): HD~16555 CC\#1 which is a true `in-between' case (see Fig.~A1), and HD~43940 CC\#1, for the reasons developped in Sect.~\ref{sec:HD43940}. The 13 other imaged point sources (32\%) have only been observed once; their status is set to `(U)ndefined' in Table~\ref{tab:cc}.

\subsection{Detection limits}
\label{sec:limdet}
For each image, we derive 6-$\sigma$ detection limits in the form of a 2-dimensional map. The detection limits are calculated by measuring for each pixel the noise (standard deviation) in a $5\times5$ box centered on that pixel (see also Lagrange \etal\ 2009c). This operation is performed with the \texttt{IDL} routine \texttt{IMAGE\_VARIANCE} (by M.~Downing). The coronagraphic images are normalized to the exposure time of the associated direct image before their detection limits are calculated; the neutral density filter transmission is also taken into account. The classic 1-dimensional detection limits are then radially extracted from the 2D maps: the 1D detection limit at a separation $\rho$ from the star position at $(\Delta \alpha,\Delta \delta) = (0,0)$ is the azimuthal median of all pixels in an annulus of mean radius $\rho$ with 5-pixel thickness. 

These 1D detection limits can be used for determining the overall properties of our survey and comparing it to other surveys. However, since the contrast is mainly limited by the presence of speckles at close separations, we believe that the 2D maps are more accurate than the 1D limits and should rather be used when refering to particular objects in the survey. At large separations from the central stars, the contrast is rather readout-noise limited, and both 1D or 2D limits could be used. 

\begin{figure*}
\begin{center}
\includegraphics[width=18cm]{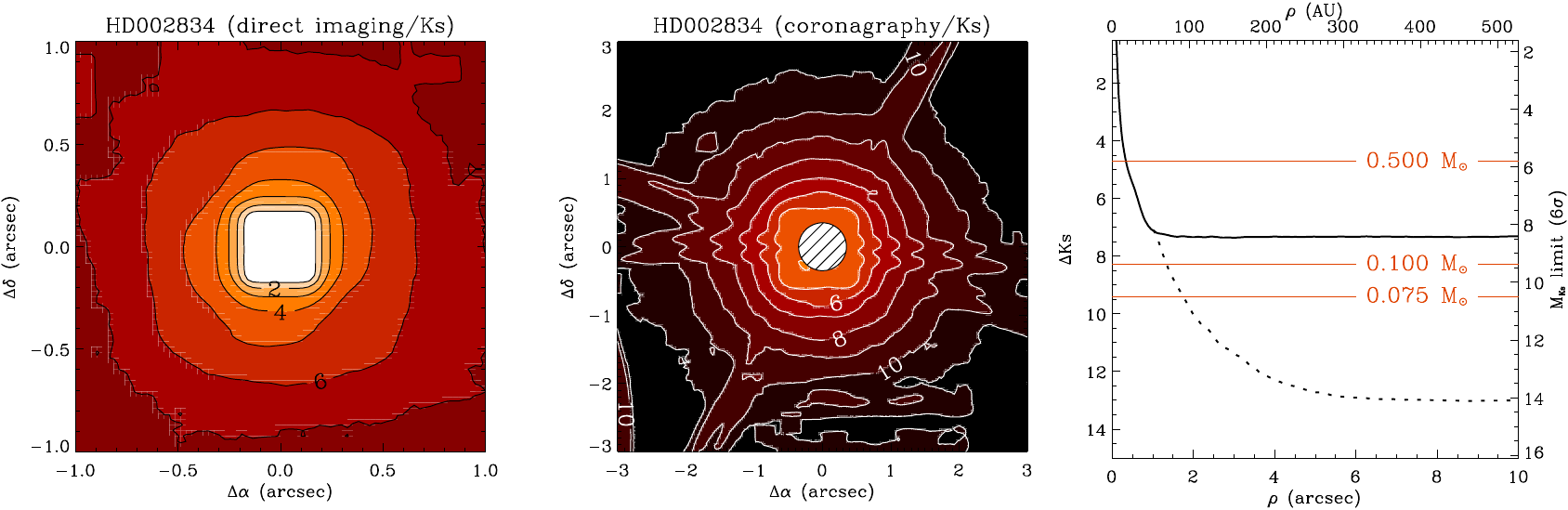}
\caption{\label{fig:A0limdet}Example of 6-$\sigma$ detection limits reached for a $K=4.7$~mag A0V star at 52.7~pc (HD~2834) using direct imaging (left panel) and coronagraphy (middle panel). Detection limits for the direct and coronagraphic image are given in $2\times2$ and $6\times6$~arcsec$^2$ fields around the star, respectively. The hatched area on the coronagraphic image represents the occulting mask with a diameter of 0\farcs7. Classic 1-dimensional 6-$\sigma$ detection limits (right panel) are extracted from the direct image detection limit map (plain line) and the coronagraphy detection limit map (dotted line) as explained in Sect.~\ref{sec:limdet}. Mass limits (red lines) are obtained by interpolating Baraffe \etal 's (1998) evolutionary model for a given stellar age (here, 700~Myr).}
\end{center}
\end{figure*}

\begin{figure*}
\begin{center}
\includegraphics[width=18cm]{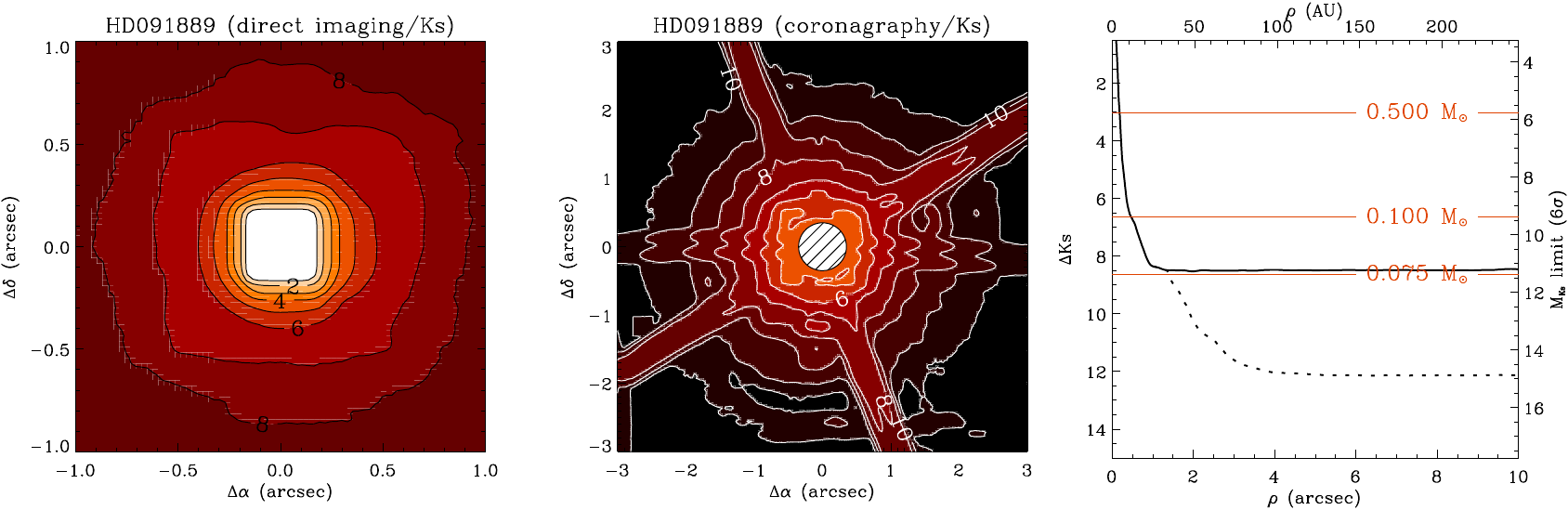}
\caption{\label{fig:F7limdet}Same as Fig.~\ref{fig:A0limdet} for a $K=4.3$~mag F7V star at 24.6~pc, and an age estimated to $\sim 4.1$~Gyr (HD~91889).}
\end{center}
\end{figure*}

\begin{figure}
\begin{center}
\resizebox{\columnwidth}{!}{\includegraphics{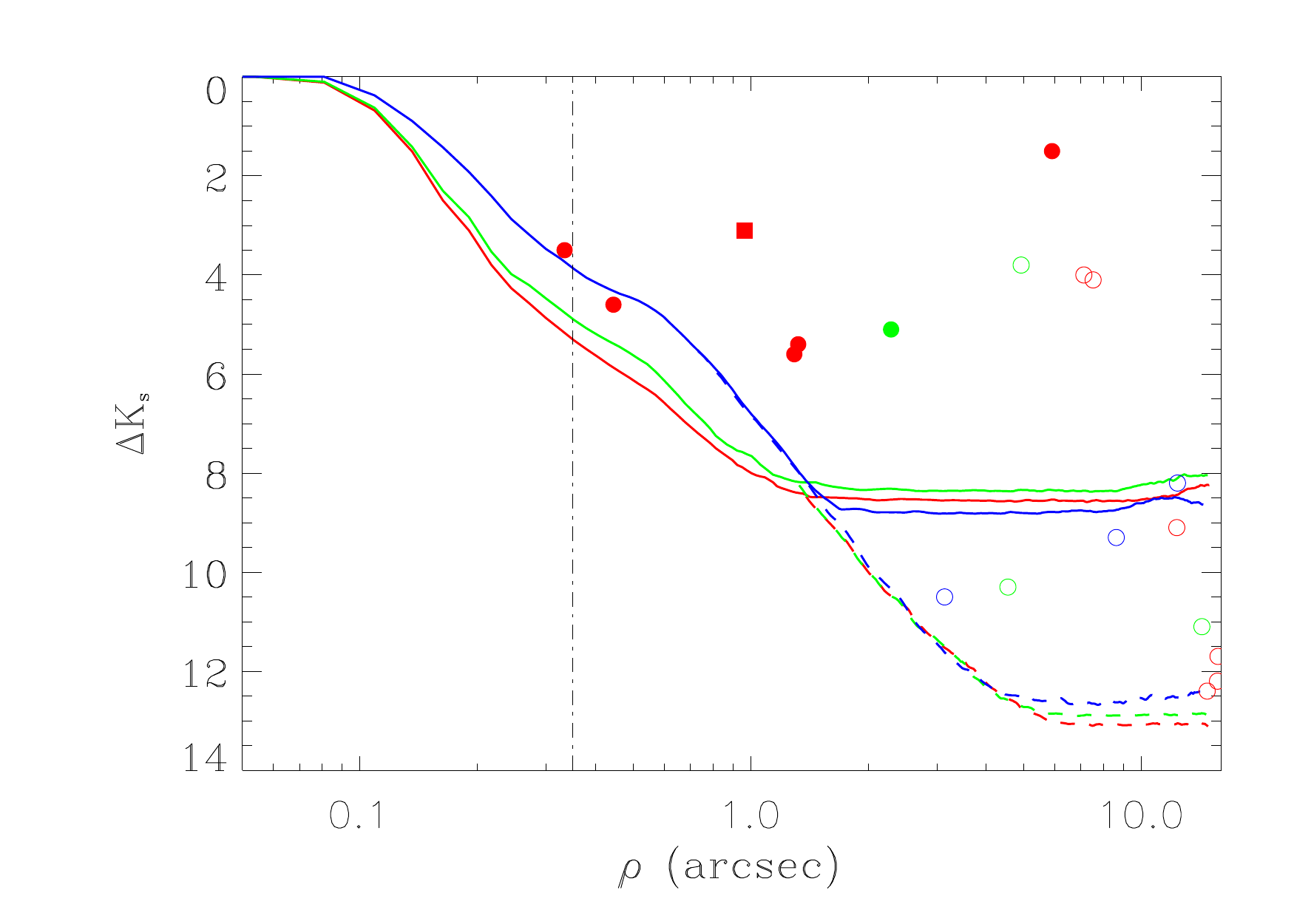}}
\caption{\label{fig:limdetK}Median 6-$\sigma$ detection limits in the \Ks\ band for F- (red), A- (green), and B-type (blue) stars, in direct imaging (plain lines) and coronagraphic (dashed lines) modes. The coronagraphic detection limits are better than the direct imaging ones from 2\arcsec\ on, for the reasons detailed in Sect.~\ref{sec:limdet}. The measured separations $\rho$ and magnitude differences $\Delta$\Ks\ of companion candidates confirmed through multi-epoch observations and those with an undefined status are reported by filled and empty circles, respectively. The confirmed companion candidate to $\iota$~Crt, which was only observed in the $H$ band, is marked by the red square assuming a $(H-K)$ colour close to 0. The dash-dotted line at $\rho = 0\farcs35$ indicates the occulting mask radius.}
\end{center}
\end{figure}

Figures~\ref{fig:A0limdet} and~\ref{fig:F7limdet}  show the 2D- and 1D-detection limits achieved both in direct imaging and coronagraphic modes for two stars from our survey: an early A-type star (HD~2834; Fig.~\ref{fig:A0limdet}) and a late F-type star (HD~91889; Fig.~\ref{fig:F7limdet}) are chosen as typical examples. Maps of the detection limits for all surveyed stars are available in Appendix~B. The overall 1D detection limits in the \Ks\ band are presented in Fig.~\ref{fig:limdetK}. Unlike the direct imaging observations, the coronagraphic imaging mode did not use a neutral density filter, and therefore enabled improved sensitivities for wider-separation (greater than 2\arcsec) sources.  With the coronagraph in place, we were able to employ large Detector Integration Times (DITs), while still avoiding detector saturation. The larger DITs helped reduce detector read-out noise, which currently limits the wide-separation sensitivities.  At separations greater than $\sim5\arcsec$, we measured $\sim4$-mag improvement between direct imaging and coronagraphic imaging modes.

Since the different spectral types of stars included in the survey have (i) different distance cut-off values (see Fig.~\ref{fig:distances}) and (ii) different median ages, it is interesting to plot the overall detection limit for each spectral type (A, F, and B) as the median of all detection limits. Note that since there are only 3 B stars in the survey, the median limit for B stars (blue curves) cannot be considered as representative as for A and F stars (red and green curves, respectively). 

In this scope, the performance in contrast is slightly better, typically by $\la 0.5$~mag, for F-type stars than for A-type stars. Using direct imaging mode, we achieve typical contrasts of 4, 6, and 8.5~mag at 0\farcs2, 0\farcs5, and $>1\arcsec$ from the star, respectively. The use of an occulting mask allows to obtain better contrast than with direct imaging from $\sim 1\farcs5$. A maximum contrast of $\sim 13$~mag is reached at separations larger than 5\arcsec\ from the star. 


\begin{figure}
\begin{center}
\resizebox{\columnwidth}{!}{\includegraphics{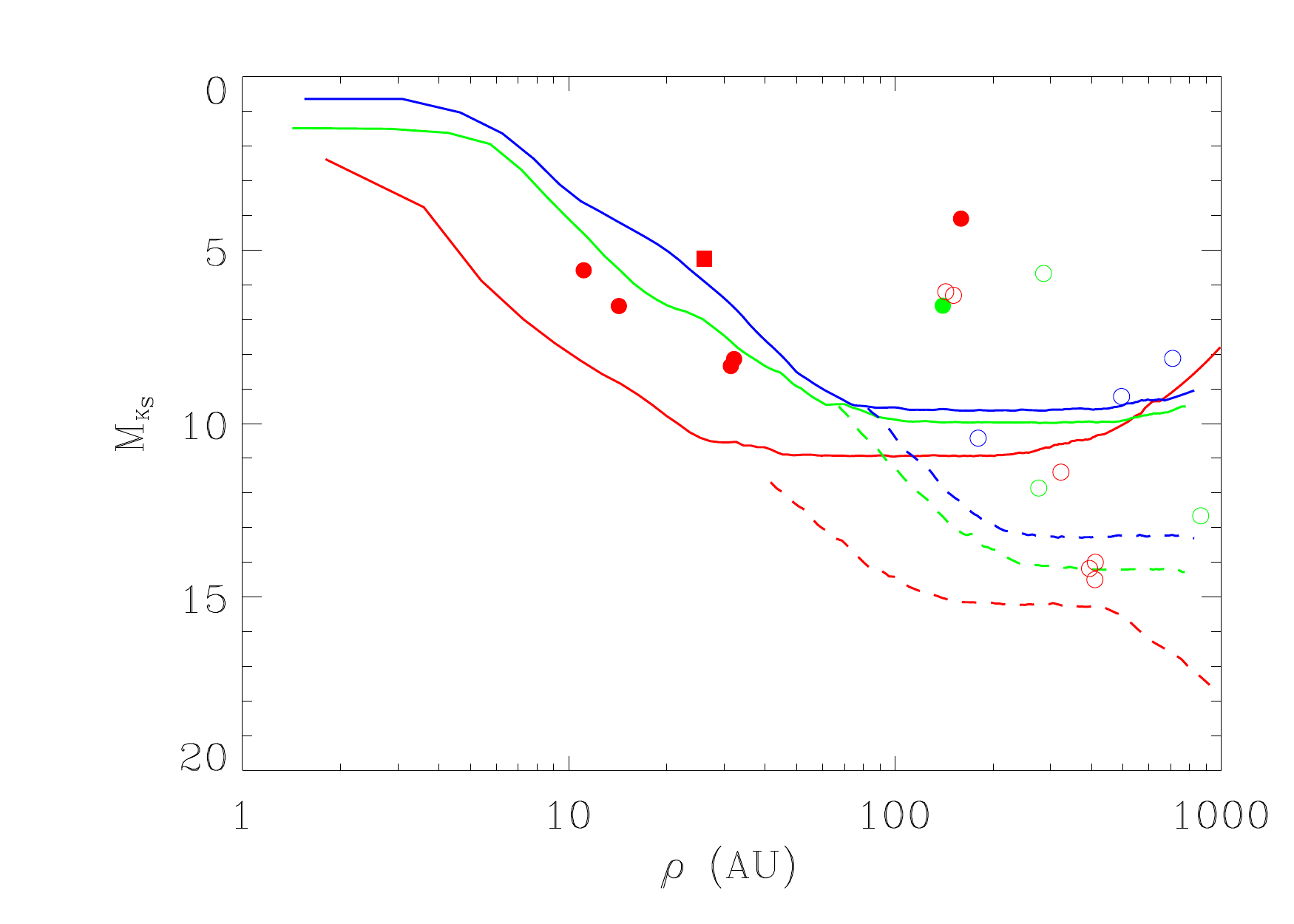}}
\caption{\label{fig:limdetAbsK}Absolute magnitudes vs.\ projected separations. The legend is the same than for Fig.~\ref{fig:limdetK}. The absolute magnitudes $M_{K_s}$ and projected separations are calculated for each object given its apparent magnitude $m_{K_s}$ in the \Ks\ band and distance $d$ from Earth. The 6-$\sigma$ detection limits shown for F- (red), A- (green), and B-type (blue) stars, in direct imaging (plain lines) and coronagraphic (dashed lines) modes are the medians of the detection limits $M_{K_s} = f(\rho(d), d, m_{K_s})$ calculated for all stars of a given spectral type.}
\end{center}
\end{figure}


\begin{figure*}
\begin{center}
\resizebox{0.45\textwidth}{!}{\includegraphics{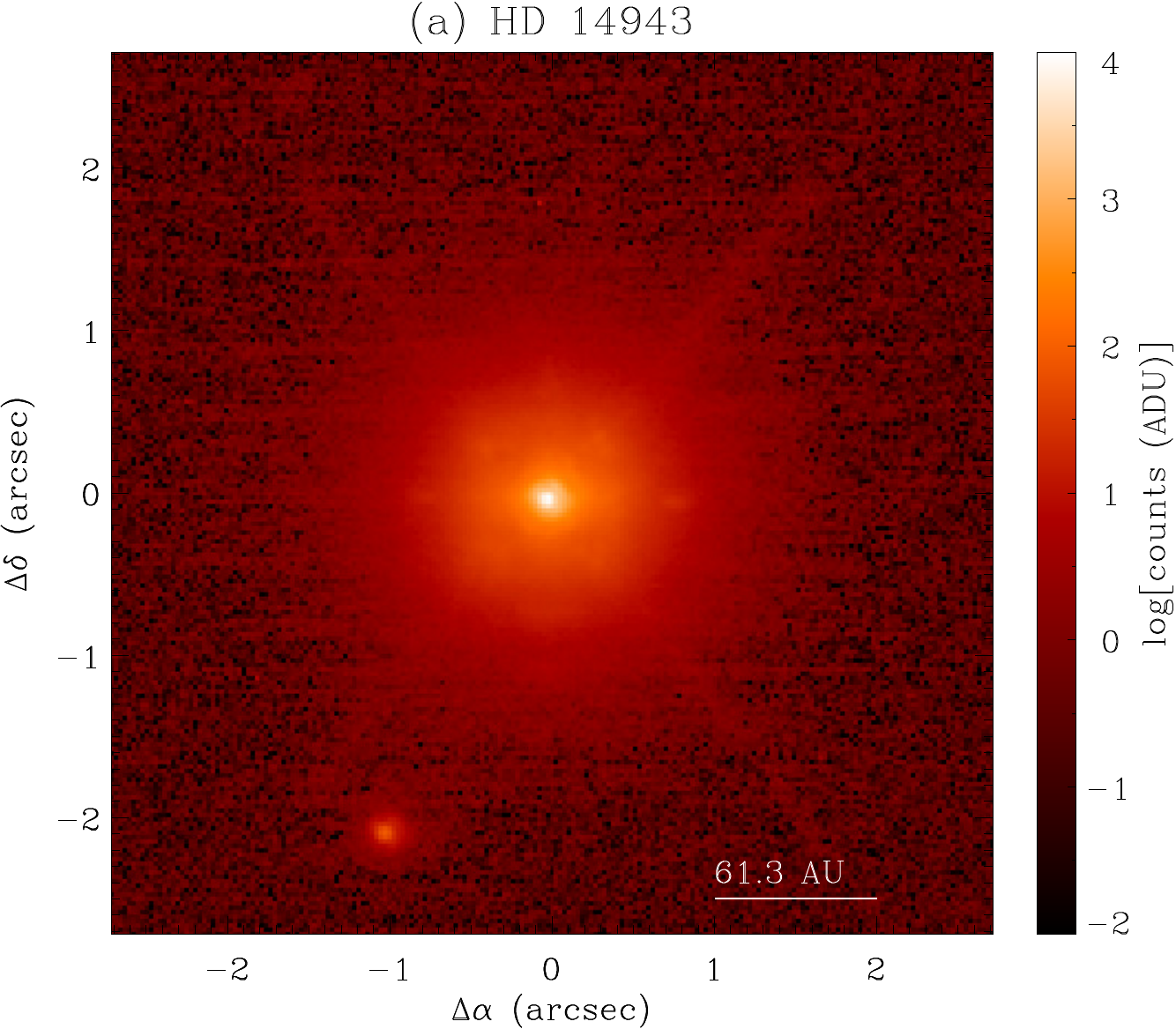}} 
\resizebox{0.45\textwidth}{!}{\includegraphics{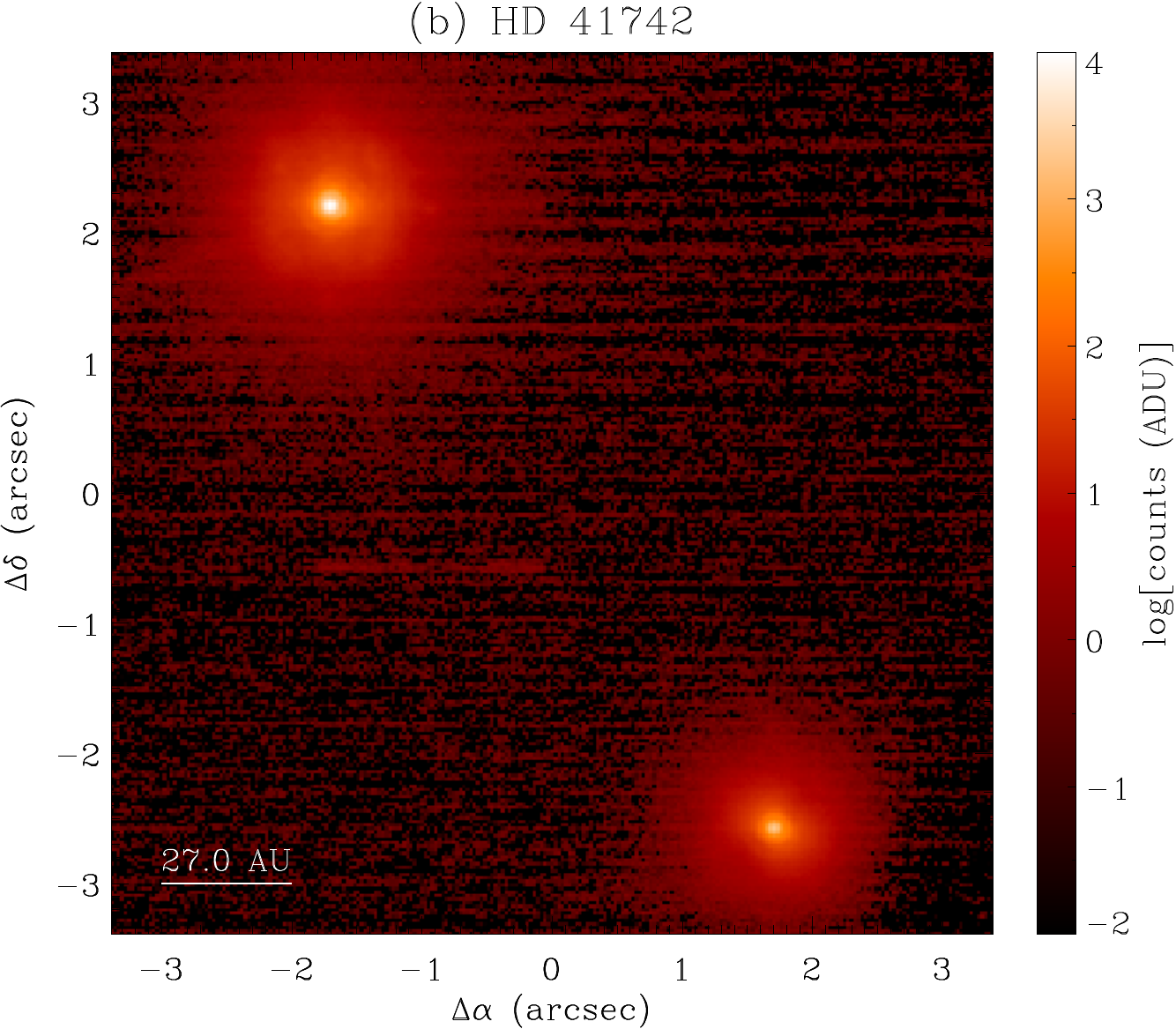}} 
\resizebox{0.45\textwidth}{!}{\includegraphics{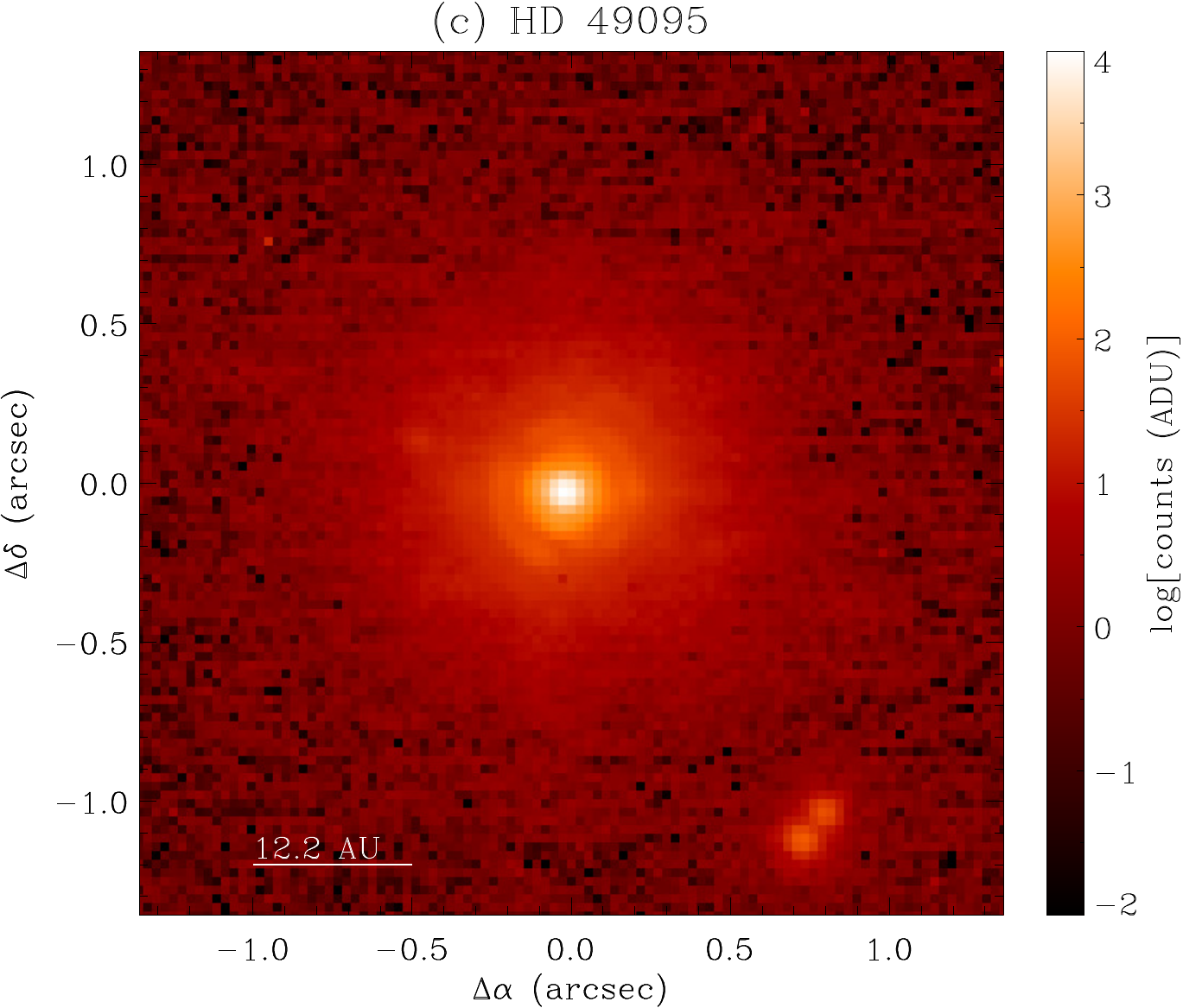}} 
\resizebox{0.45\textwidth}{!}{\includegraphics{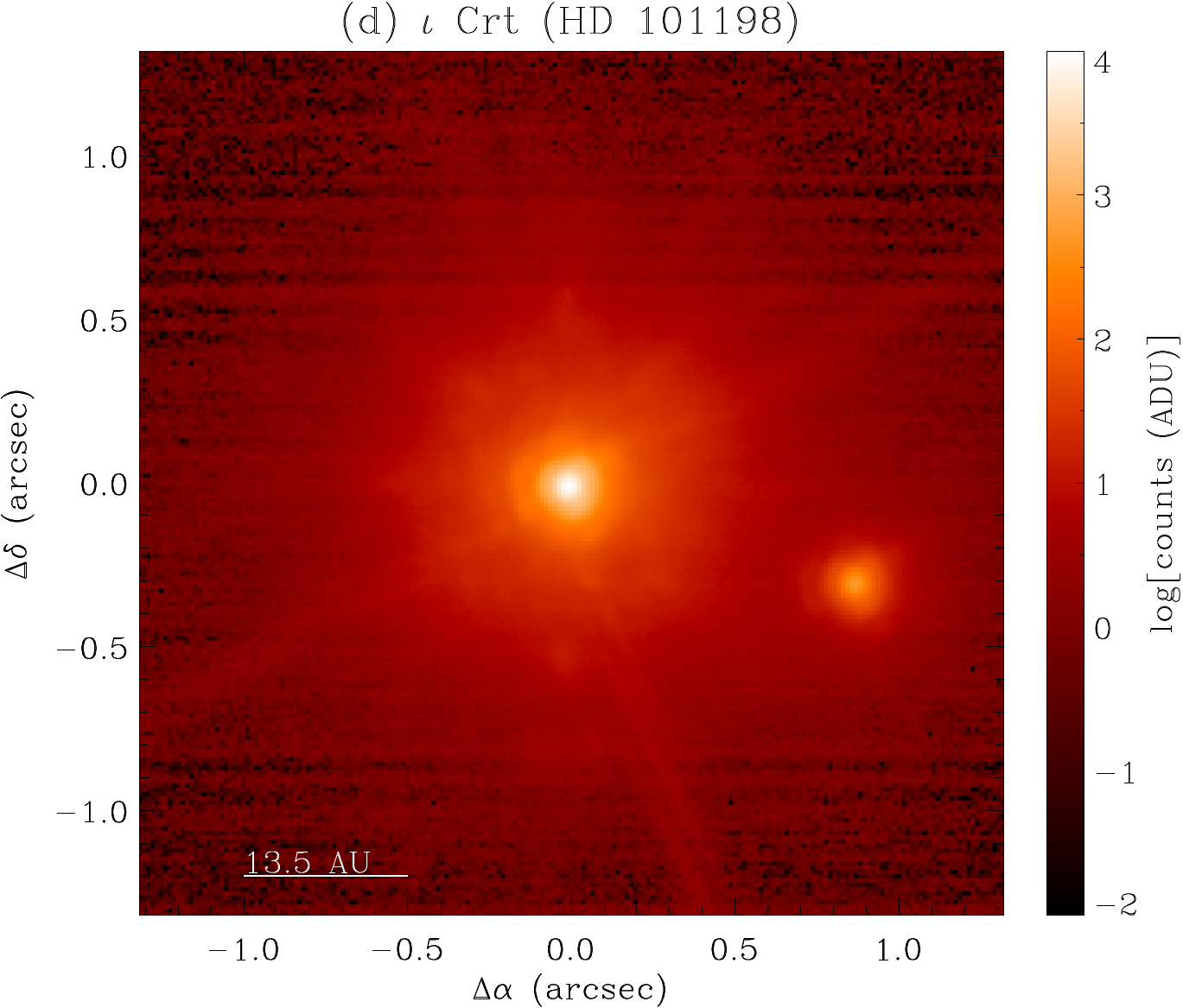}} 
\resizebox{0.45\textwidth}{!}{\includegraphics{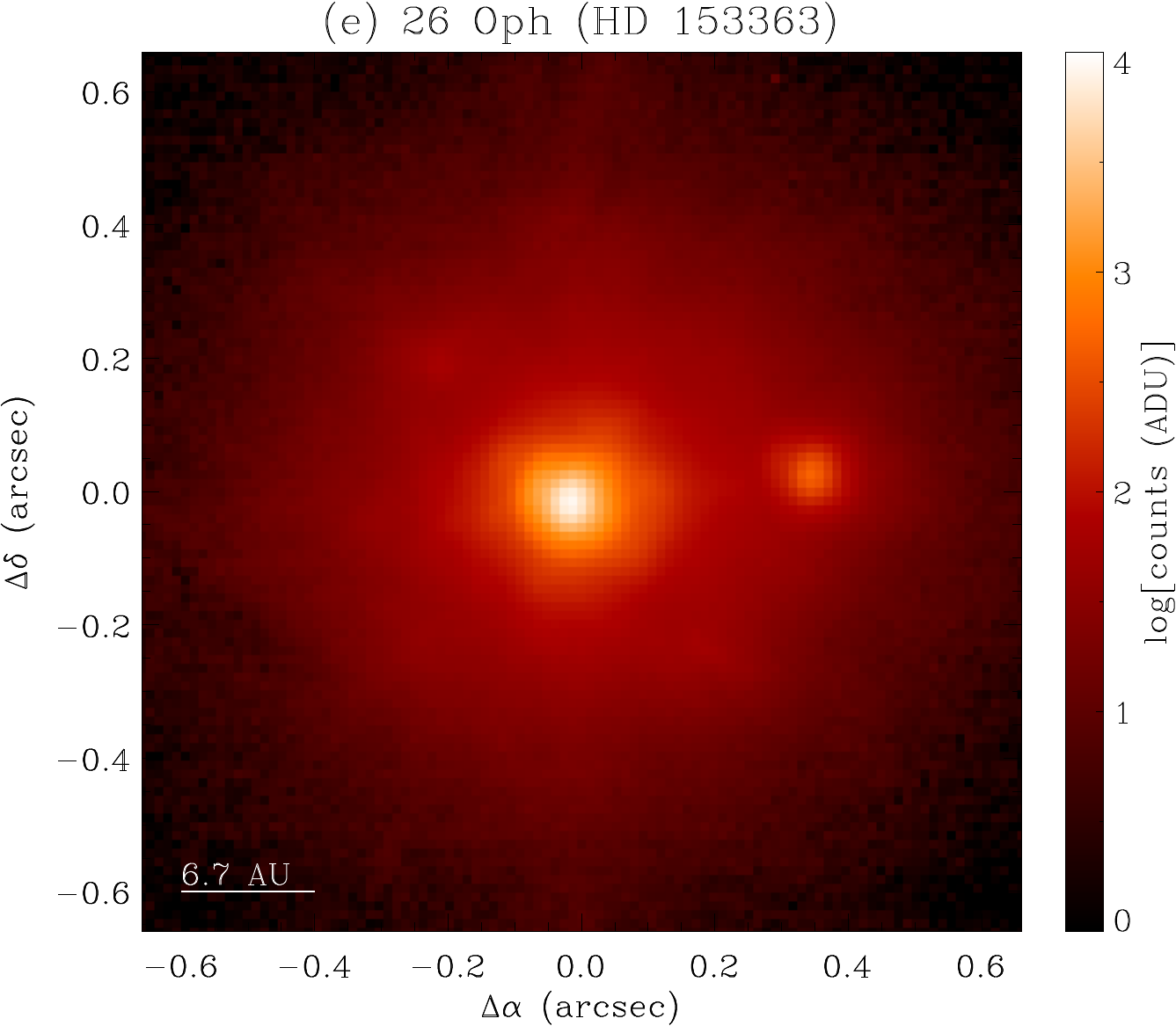}} 
\resizebox{0.45\textwidth}{!}{\includegraphics{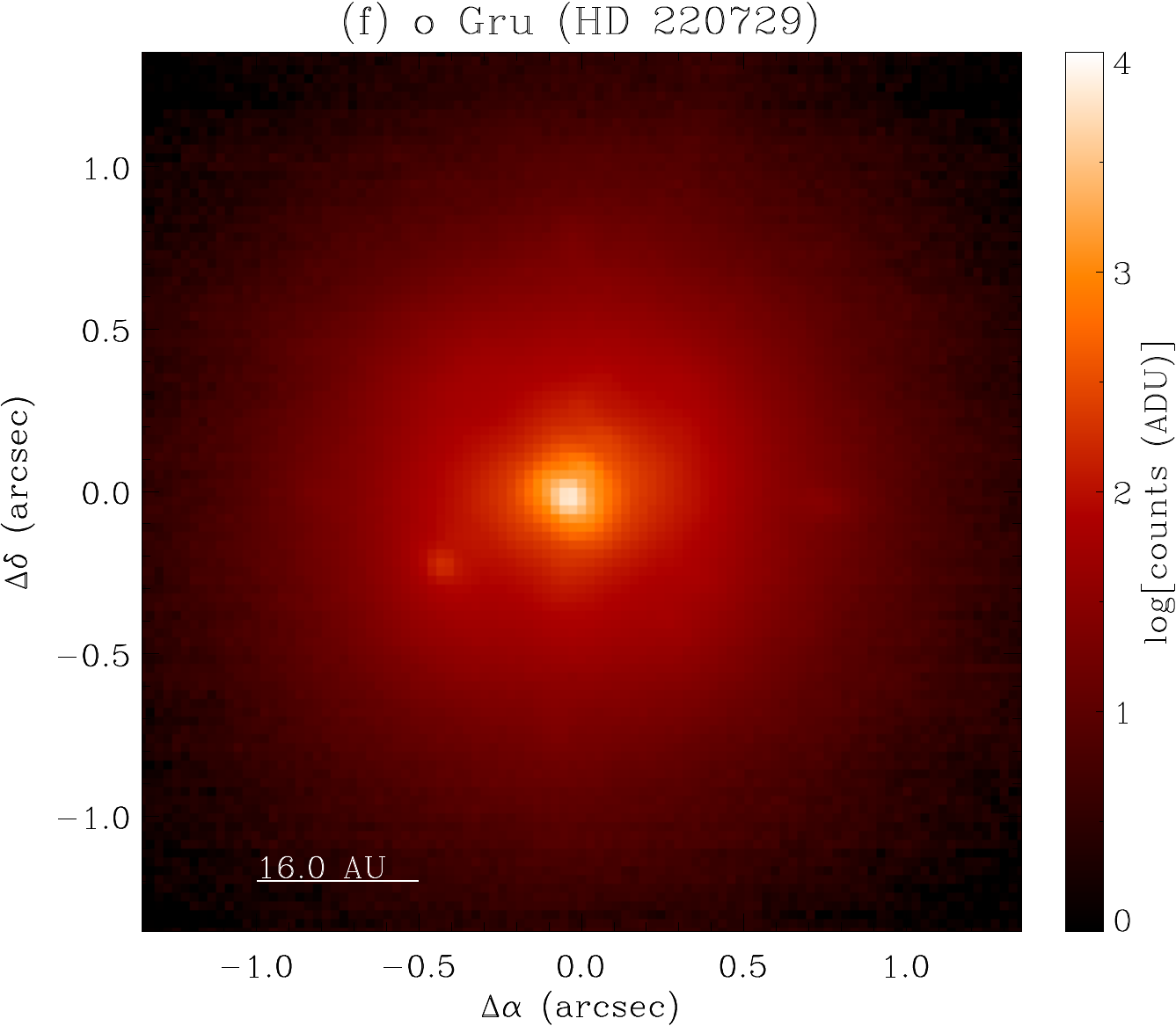}} 
\caption{\label{fig:mosaic}Stars with companion candidates confirmed through multi-epoch observations. North is up and east is to the left. Counts on the detector are displayed with a logarithmic scale.}
\end{center}
\end{figure*}


\subsection{Companion candidates detected}
Companion candidates are detected at all separations above the detection limits. Table~\ref{tab:cc} presents the observing dates, integration times, projected separations, position angles, and magnitude differences of the 41 point sources we identified throughout our observing campaign. Based on the method presented in Sect.~\ref{sec:companionship}, we were able to determine that 20 objects (49\% of sources detected) were background sources and 6 objects (15\% of sources detected) were in fact comoving with their primary stars. These confirmed companions are shown in Fig.~\ref{fig:mosaic} and their physical properties summarized in Table~\ref{tab:CCprop}. Because a second epoch observation could not have been obtained for all targets, 15 point sources are still of undefined nature. In the following, we detail the properties of all confirmed companions as well as those of a selection of remarkable cases of companions with undefinied status and some that we rejected as comoving objects.

\subsubsection{Confirmed comoving companions}
For companions comoving with their primary stars, the projected physical separation in astronomical units can be simply extracted from the knowledge of the separation measured on the sky and the parallax (usually, the \emph{Hipparcos} parallax; ESA 1997). The masses of companions can be retrieved using mass-magnitude relationships yielding from evolutionary models. The detection limits of our survey of A and F stars (Fig.~\ref{fig:limdetK}) show that we are sensitive to companions in the brown-dwarf to low-mass star regimes. Thus, we chose the evolutionary models for solar metallicity low-mass stars and the associated mass-magnitude relationships provided by Baraffe \etal\ (1998) to determine the masses of our detected comoving companions. Pratically, the masses were obtained for given absolute magnitudes and ages by interpolations into Baraffe \etal 's tables (VizieR On-line Data CatalogueJ/A+A/337/403).

\paragraph{\emph{HD~14943---}}
A companion with an absolute magnitude of $M_K=6.6$~mag has been identified at two different epochs $2\farcs3$ (141~\au) away from the A5V star HD~14943 (HIP~11102). According to Baraffe \etal 's (1998) evolutionary models for solar-metallicity low-mass stars, this companion should be a 0.4~\Msun\ star, assuming an age of 850~Myr, determined in the frame of Su \etal 's (2006) survey for debris discs around A stars with \emph{Spitzer}/MIPS. Hence, HD~14943B, shown in Fig.~\ref{fig:mosaic}a, could be a main sequence M2 star (Cox 2000).\footnote{According to Drilling \& Landolt's calibration of $MK$ spectral types (in Cox 2000, p.~389, Table~15.8). This star exhibits radial velocity variations (Lagrange \etal\ 2009).}

\paragraph{\emph{HD~41742---}} 
This high proper-motion star is part of a physical ternary hierarchical system (Multiple Star Catalogue --- MSC ---, Vizier Online Data Catalogue J/A+AS/124/75, Tokovinin 1997). We identify CC1 to component B (CCDM~J06046-4504B; see Fig.~\ref{fig:mosaic}b), a 0.79~\Msun\ star with a period of $\sim 1\,400$~yr. The mass given in the MSC was determined from the measured $B-V=1.01$, typical of a K0/K2 star and according to Cox (2000). The two other companion candidates (CC2 and CC3) are background objects. The component C of the system catalogued in the MSC is too far to enter NaCo or PUEO field-of-view ($\rm \rho_C=196\farcs7$). Furthermore, the primary star was shown to be a spectroscopic binary according to the radial velocity curves obtained by (Lagrange \etal\ 2009). The imaged companion is not responsible for the radial velocity variations; hence, HD~41742 is a member of a quadruple system. We measured for HD~41742B (CC1) a projected separation of 5\farcs9, i.e., 159~\au\ and an absolute magnitude $M_K = 4.0$~mag, which translates into a 0.8-\Msun\ star using Baraffe \etal 's (1998) mass-magnitude relationships, in good agreement with the mass previously determined. This estimation was made assuming an age of 3.7~Gyr, according to the Geneva-Copenhagen Survey of Solar Neighbourhood (GCSSN, Vizier Online Data Catalogue V/130, Holmberg, Nordstr\"om \& Andersen 2009).

\begin{figure}
\begin{center}
\resizebox{\columnwidth}{!}{\includegraphics{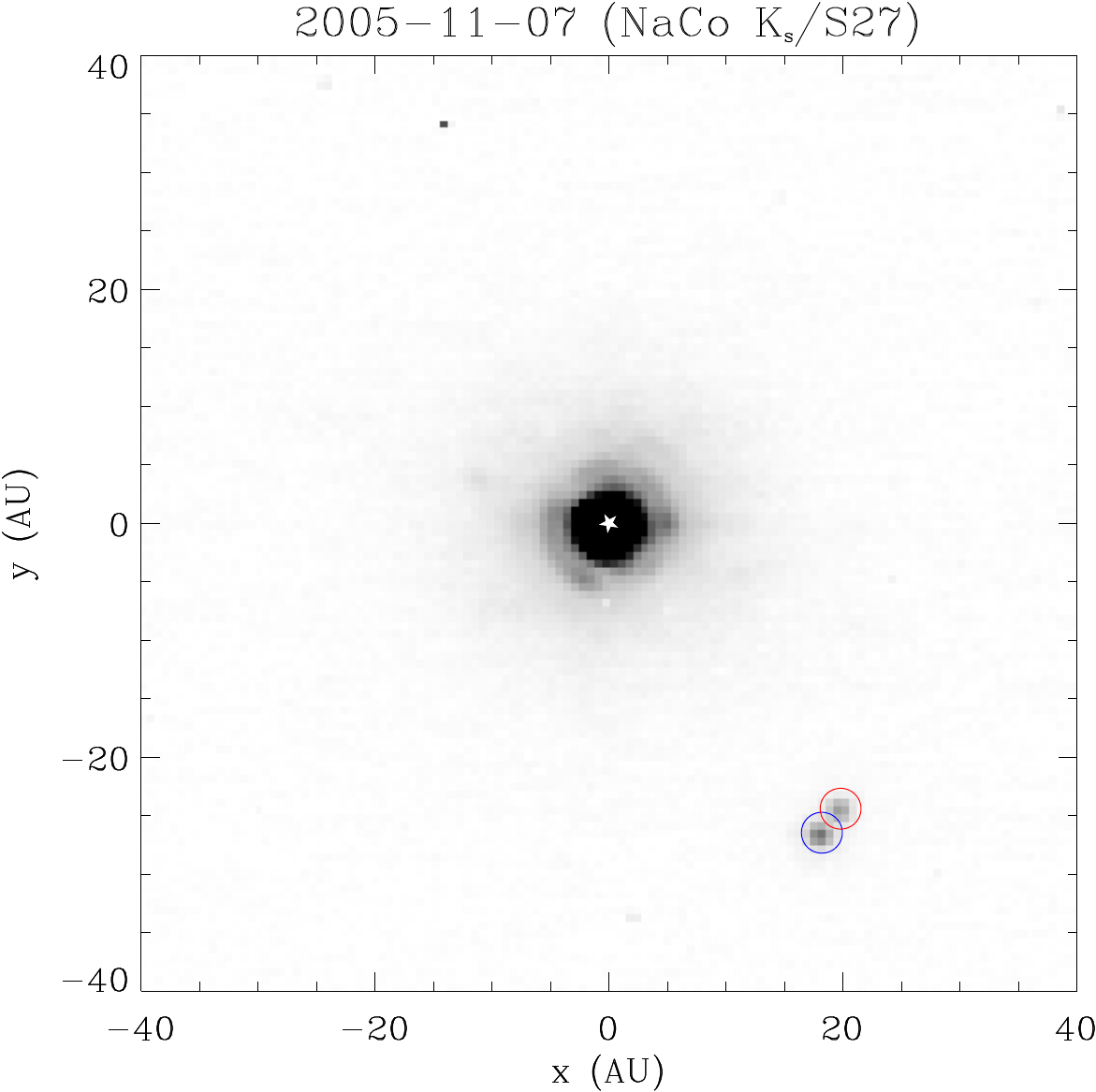}}
\caption{\label{fig:HD49095_id}The F6.5V star HD~49095 and its binary comoving companion composed of CC1 (blue circle) and CC2 (red circle). The axis are labelled in projected separation with respect to the primary star centroid position (white star). The positions of objects on this image are retrieved using the deconvolution algorithm described in Sect.~\ref{sec:results:extraction}.}
\end{center}
\end{figure}

\begin{figure}
\begin{center}
\resizebox{\columnwidth}{!}{\includegraphics{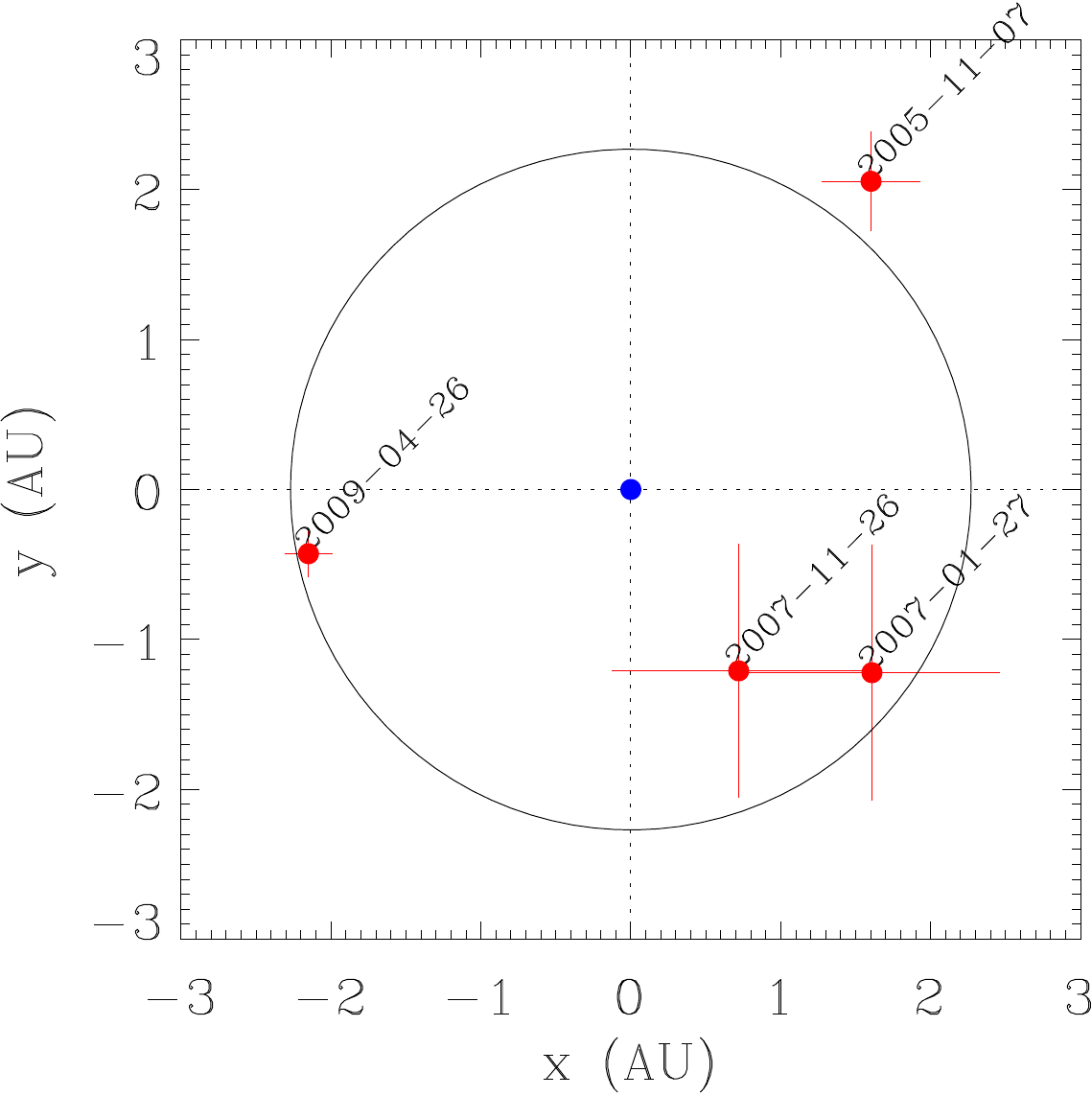}}
\caption{\label{fig:HD49095_orbit}Motion of the binary companion CC2 of HD~49095 (zoom from Fig.~\ref{fig:HD49095_id}) relatively to CC1 (blue point) between 4 different epochs covering 3.5~yr. The best-fit circular and face-on orbit is figured, with a semi-major axis of 2.27~\au.}
\end{center}
\end{figure}

\paragraph{\emph{HD~49095---}}
This high proper-motion star with radial velocity variations (Lagrange \etal\ 2009) is referenced as a binary by Makarov \& Kaplan (2005) based on the difference of proper motions measured between two different catalogues (`$\Delta \mu$ binary'), namely, the Hipparcos and Tycho-2 catalogues. Here, we image the binary component and resolve it as a tight binary in which orbital motion is detected over several epochs. The system is shown as seen on 2005-11-07 in the \Ks\ band with NaCo in Figs.~\ref{fig:mosaic}c and~\ref{fig:HD49095_id} and the revolution of the B (CC1) and C (CC2) components around their center of mass is shown over 3.5 years (4 epochs) in Fig.~C1. The position of HD~49095C with respect to HD~49095B was retrieved using the deconvolution algorithm described in Sect.~\ref{sec:results:extraction} and is plotted in Fig.~\ref{fig:HD49095_orbit}. For the two VLT images (taken on 2005-11-07 and 2009-04-26) where the HD~49095BC system is well resolved, we assumed 1-$\sigma$ position errors of 0.25 pixel. A position error of 0.5 pixel has been assumed for CFHT images (taken on 2007-01-27 and 2007-11-26) where the system is barely resolved. The tight system projected separation is seen to vary between 1.4 and 2.6~\au. Yet, because we do not have enough points to fit a precise orbit, we simply calculated a best-fit circular orbit, assuming the system is seen face-on, and determined a semi-major axis of 2.27~\au\ ($\chi^2 = 3.3$ with 1 degree of freedom). HD~49095C accomplishes roughly 60\% of this circular orbit in 3.5~yr, which leads to a revolution period of 5.7~yr. Using Kepler's third law, we can then determine a total mass for the BC system of 0.11~\Msun. This is about three times smaller than the total mass determined photometrically using Baraffe \etal 's (1998) models ($0.175 + 0.13 = 0.305$~\Msun) given the measured magnitudes in the \H\ band of $M_{HB} = 8.5$ and $M_{HC} = 9.0$~mag and the age of 2~Gyr provided by the GCSSN. We attribute this discrepancy to the fact that the orbit is not well constrained with 4 epochs. 


\paragraph{\emph{HD~101198---}}
The $\iota$~Crateris (HD~101198; CCDM~J11387-1312A) entry in the CCDM mentions a companion (KUI~58B) at $\rho=1\farcs4$ and $\theta=-134\degr$ in 1934, with a visual magnitude $V = 11$. The star is recorded as a probable single star in the Double and Multiple Systems Annex of the Hipparcos Catalogue (ESA 1997), yet Makarov \& Kaplan (2005) as well as Frankowski, Jancart \& Jorissen (2007) see it as a $\Delta \mu$ binary. Were KUI~58B a background contaminant, it would lie in 2009 at $\rho \approx 13\farcs45$ and $\theta \approx -141\deg$, so that we should be able to detect it on NaCo \Ks\ images, unless its $V-K$ colour index is below $\sim -1.5$. The other possibility is that KUI~58B is indeed bound to $\iota$~Crt. In this case, it is likely that CC1, shown in Fig.~\ref{fig:mosaic}d, is that object. This would imply for CC1 colour indexes $V-K \approx 11-7.25 = 3.75$ and $V-H \approx 11-7.44 = 3.56$, typical of a M3 dwarf (Ducati \etal\ 2001)\footnote{Here, we assume that the binary has the same age as the primary ($\sim 5$~Gyr) and is therefore on the main sequence.}. The mass derived from the measured $M_H = 5.2$~mag using evolutionary model and assuming an age of 4.8~Gyr (from the GCSSN) is 0.57~\Msun, more typical of a M0 or a late K star (Cox 2000). Furthermore, the primary shows radial velocity variations (Lagrange \etal\ 2009).

\paragraph{\emph{HD~153363---}}
26~Ophiuchi (HD~153363) is referenced as a radial-velocity variable (Lagrange \etal\ 2009) as well as an astrometric binary (Proper motion derivatives of binaries, Vizier Online Data Catalogue J/AJ/129/2420, Makarov \& Kaplan 2005; HIP binaries with radial velocities, Vizier Online Data Catalogue J/A+A/464/377, Frankowski, Jancart \& Jorissen 2007), yet no orbit information is available in the litterature. The binary component expected by astrometric measurements could well be CC1. In fact, we determined that the motion of CC1 seen between August 2008 and April 2009 cannot be due to the combination of the stellar proper motion and the parallactic motion, which are for this low-proper motion star of similar amplitudes. It can be seen in Fig.~A1 that the position of CC1 in April 2009, imaged in Fig.~\ref{fig:mosaic}e, is not compatible with that of a background field object. Hence, the motion of CC1 can rather be explained by the orbital motion around 26~Oph. We observed 26~Oph~B in three band passes and measured $\Delta K_s = 2.1$, $\Delta H = 2.7$, and $\Delta J = 3.1$. Given the 2MASS $JHK$ magnitudes of 26~Oph and its distance of 33.3~pc, the companion has absolute magnitudes $M_{K_s}=4.2$, $M_H=4.9$, and $M_J=5.5$. These magnitudes are compatible, within the given uncertainty range of $\pm0.4$~mag, with a 1.0--1.3-Gyr-old 0.7-\Msun\ object according to Baraffe \etal 's (1998) model. Assuming the $J-K_s$ and $H-K_s$ colours are each overestimated by $\sim0.4$~mag (a 1-$\sigma$ variation), they indeed correspond to the colours of an early K dwarf (Ducati \etal\ 2001). Additional fainter and farther components of this system may exist; however, the confirmation of their companionship would require a second epoch with coronagraphic observation. Meanwhile, we can already exclude CC2, also seen on the direct images, as an additional physical companion to 26~Oph. 

\paragraph{\emph{HD~220729---}}
Lagrange \etal\ (2009a) identified $o$~Gruis (HD~220729) as a spectroscopic binary and noted that it was associated to a ROSAT source by Suchkov, Makarov \& Voges (2003). In this work, we confirm the binary nature of this star as we resolve it as a tight (0\farcs4 or 14.9~\au) and contrasted ($\Delta K = 4.7$) binary system, as can be seen in Fig.~\ref{fig:mosaic}f. We estimate the absolute magnitude of \object{$o$~Gru~B} to $M_K=6.7$ for an age of 1.1~Gyr (GCSSN), which yields a mass of 0.3~\Msun, i.e., a M3/M4 main sequence star. The companion detected in the adaptive-optics survey is not responsible for the radial-velocity shifts, hence $o$~Gru is a triple system.

\begin{table}[!t]
\begin{center}
\caption{\label{tab:CCprop}Physical properties of comoving companions.}
\begin{tabular}{rlccc}
\hline \hline
\#& Star      & CC\# & Projected Separation & Mass    \\
  &           &      & (\au)                & (\Msun) \\
\hline
5 & HD 14943  & 1    & 141        & 0.40    \\
15& HD 41742  & 1    & 159        & 0.80    \\
17& HD 49095  & 1    & 32.3       & 0.175   \\
17& HD 49095  & 2    & 31.9       & 0.13    \\
22& HD 101198 & 1    & 25.0       & 0.57    \\
25& HD 153363 & 1    & 11.3       & 0.70     \\
36& HD 220729 & 1    & 14.9       & 0.30    \\
\hline
\end{tabular}
\end{center}
\end{table}

\subsubsection{Companion candidates with undefined or ambiguous status}

\paragraph{\emph{HD~32743---}}
$\eta^1$~Pictoris (HD~32743, HIP~23482) is catalogued in the CCDM as a binary with an astrometry recorded in 1946 of $\rho = 10\farcs6$ and $\theta = -162\degr$, and a photometry of $\Delta V = 7.4$~mag (no associated uncertainties on these parameters are reported). We observed this object during a single epoch on 2005-11-07, and obtained the following astrometry for CC1: $\rho=12\farcs212 \pm 0\farcs0.0285$ and $\theta=-176.32\pm0.12\degr$. We measured infrared apparent magnitudes of $J=12.6$ ($\Delta J=7.8$) and $K_s=11.8$ ($\Delta K_s = 7.4$) for CC1, which seem compatible with the visible apparent magnitude ($V=13$) of the known companion to $\eta^1$~Pic. Hence, it is likely that CC1 is the binary companion to $\eta^1$~Pic and catalogued as \object{CD-49~1541B}. To estimate whether CC1 is comoving with $\eta^1$~Pic, we produced a proper motion plot similar to those of Fig.~A1 and that is presented in Fig.~D1. We assumed arbitrarily large error bars on the 1946 astrometric parameters tabulated in the CCDM. The position of the object measured in 2005 corresponds to that of a background object, non-comoving with the primary star. We obtain $P_\mathrm{bkg} = 24.9\%$. The lack of reported uncertainties in the CCDM does not allow to firmly conclude on the companionship status of CC1/CD-49~1541B; we consider it as a probable background contaminant. We are nevertheless resolving it as a tight apparent binary and, since a second tight binary system is also present in the surroundings of $\eta^1$~Pic, the system will need to be monitored in the future.

\paragraph{\emph{HD~43940---}}\label{sec:HD43940}
Clearly seen in 2005-November observations in the \Ks\ band, the close companion candidate to HD~43940 ($\rho = 0\farcs22$) is harder to resolve from the primary star in 2008-August, both in \Ks\ and \H\ bands. This may be due, however, to the poorer observing conditions compared to 2005-November. A final attempt to characterize this new tight system was done on 2009-April, but the companion could barely be resolved from the primary star in the \J\ band. HD~43940 has very low proper motions in the direction of CC1 ($\delta_\alpha = -5.51$~mas~yr$^{-1}$, $\delta_\delta=-22.63$~mas~yr$^{-1}$), but since the separation is small, it is hard to disentangle between the two possibilities: (i) the system is apparent, CC1 is a nearly motionless background contaminant apparently getting closer to HD~43940 because of this star proper motions, or (ii) the system is physically bound and the difficulty to resolve it as time passes results from the orbital motion of CC1 around the primary. 

\paragraph{\emph{HD~158094---}}
The B8 star $\delta$~Ar\ae\ (HD~158094) is part of an optical (Torres 1986) double system (CCDM~17311-6041) whose secondary component lies out the field of view ($\rho=47\farcs4$, $\theta=313\degr$). Concerning the star itself, \emph{Spitzer}/MIPS 24-\micron\ photometry indicates an age of 125~Myr and nonexistent infrared excess (Rieke \etal\ 2005). Low-mass companions were searched for around this star by Hubrig \etal\ (2001) using the AO system ADONIS on the ESO 3.6-m telescope in La Silla. These authors report detection limits of $\Delta K \sim 5$~mag and $\Delta K \sim 9$ for separations of $\sim 2\arcsec$--$5\arcsec$ and $\ga 4\arcsec$, respectively. This means CC3 ($\rho = 12\farcs3587$, $\theta=62.0\degr$, $\Delta K = 8.2$) could have been detected by Hubrig \etal\ (2001): it is slightly above their detection limit, and close to the edge of, yet within, their $24\arcsec \times 24\arcsec$ field of view, centered on the star, in 1999-March. Note that if CC3 were a background contaminant, then given the star proper motion, it would have been closer to star in 1999 ($\rho = 11\farcs3909$).

\subsection{Notes on some remarkable non-comoving multiple systems}

\paragraph{\emph{HD~3003---}}
$\beta^3$~Tucan\ae\ (HD~3003) was previously known as a multiple star (CCDM~J00327-6302AB). Were this system comoving, we should be able to retrieve both components, separated by 6\farcs4 in 1925 (as tabulated in the CCDM). Only one star is visible in the 2005 NaCo image, hence this system is only a visual assocation.

\paragraph{\emph{HD~50445---}}
The star HD~50445 is not catalogued as being part of a multiple apparent or physical system. However, it lies at a relatively low Galactic declination ($b=-15.6\degr$) and at least 5 point sources have been detected in 2005-November in the 27\arcsec -wide field of NaCo S27 camera, using the coronagraphic mask. All 5 objects were observed again in 2009-April; a software issue during the observations unfortunately prevented recording a direct image of the primary star at this epoch, hence the astrometric precision is lower for the 2009 observations (typically 1~pixel, i.e., 30~mas). We are nevertheless confident that all candidate companions are background contaminants. Note that CC2 can be seen in the infrared Digitalized Sky Survey 2 image taken on 1985-December at $\rho \sim 14\arcsec$ and $\theta \sim 155\degr$ (the apparent motion of the background contaminants on the sky between 1985 and 2005 is $\mu_\alpha \sim +0\farcs8$ and $\mu_\delta \sim +1\farcs2$). 

\paragraph{\emph{HD~177756---}}
$\lambda$~Aquil\ae\ (HD~177756) is one of the B stars included in the Hubrig \etal\ (2001) survey. These authors did not report any companions in the $K$ band with a magnitude difference up to $\Delta K \sim 9$. The observed CC1 has a magnitude difference of $\Delta K = 4$. It should have been seen in the 1999 ADONIS image, as this background contaminant was closer to the star at this epoch, unless the field of view of $24\arcsec \times 24\arcsec$ reported by Hubrig \etal\ is not squarred (depending on the dithering pattern used by these authors).

\paragraph{\emph{HD~224392---}}
Members of the Tucana association, including the Vega-like star $\eta$~Tucan\ae\ (HD~224392; see Mannings \& Barlow 1998 for the infrared excess), were surveyed in the infrared by Neuh\"auser \etal\ (2003) with the SHARP-I infrared imager on the New Technology Telescope (NTT) at La Silla. These authors report detection limits in the K band of $\Delta K = 8.3$ at $>100$~\au\ from the star. However, they could not have detected CC1 for it was out of SHARP-I $13\arcsec\times13\arcsec$ field-of-view. Zuckerman, Song \& Webb (2001) note that as part of the IRAS Faint Source Catalogue (Vizier online data catalogue II/156A; Moshir 1989), $\eta$~Tuc might also be a multiple star, yet in this case we do not resolve it. Besides, being a HARPS constant (Lagrange \etal\ 2009a) makes it unlikely to be a spectroscopic binary. We do detect two companion candidates far from the star, which are not comoving with it.

\section{Discussion}
\label{sec:discussion}

\subsection{Multiplicity of early-type stars in the sample}
We found that 6 stars out of 38 in the sample have confirmed low-mass companions, yielding an observed multiplicity fraction of 16\% for our sample. This number can be seen as a lower limit value for the sample (i) because we have imaged companion candidates around 7 other stars in the sample, without having the possibility to determine whether they are true companions or background objects, (ii) because we could have missed close binaries below our detection limits, and/or (iii) because some wide-separation binary components could have been too close to the line of sight to be resolved at the time of observations. It is also certainly a lower-limit value regarding the global multiplicity fraction of early-type stars, since our survey is biased against spectral binaries or visual binaries with companions at $\leq5\arcsec$, as described in Sect.~\ref{sec:biases}. 

In the case all the undefined candidates are bound to their primary stars, the multiplicity fraction of sample stars could be as high as 34\%. More specifically, the multiplicity fraction for F stars in the sample is $5/16\approx31\%$; it is $1/22\approx5\%$ for A+B stars. If we assume that unconfirmed companions are all physically bound companions, then these fractions would rise to 50\% and 23\% for F and A+B stars, respectively. The multiplicity fraction could be even higher than these figures due to narrow-separation binaries and orbital projection effects.

Considering confirmed companions only, it would seem surprising that the multiplicity fraction of F stars is 6 times larger than that of A stars, whereas it is usually expected that the multiplicity fraction increase with the stellar mass. Actually, this is a bias that can be attributed to the fact that we are probing different physical separations for different stellar types. In order to have about the same number of F and A stars included in this volume-limited survey, we have to go farther to include as many A stars as F stars: the median distance of sample F stars is 27.0~pc while it is 54.9~pc for sample A and B stars (see Fig.~\ref{fig:distances}). Thus, we are probing smaller physical separations for F stars than for A and B stars, as shown in Fig.~\ref{fig:limdetK}b. Five out of 6 confirmed companions detected next to F stars are located between 10 and 40~\au, at projected separations not or hardly reached for the A stars in the sample. On the contrary, a confirmed companion located at $>100$~\au\ was found for each type of stars. 

\begin{figure}
\begin{center}
\resizebox{\columnwidth}{!}{\includegraphics{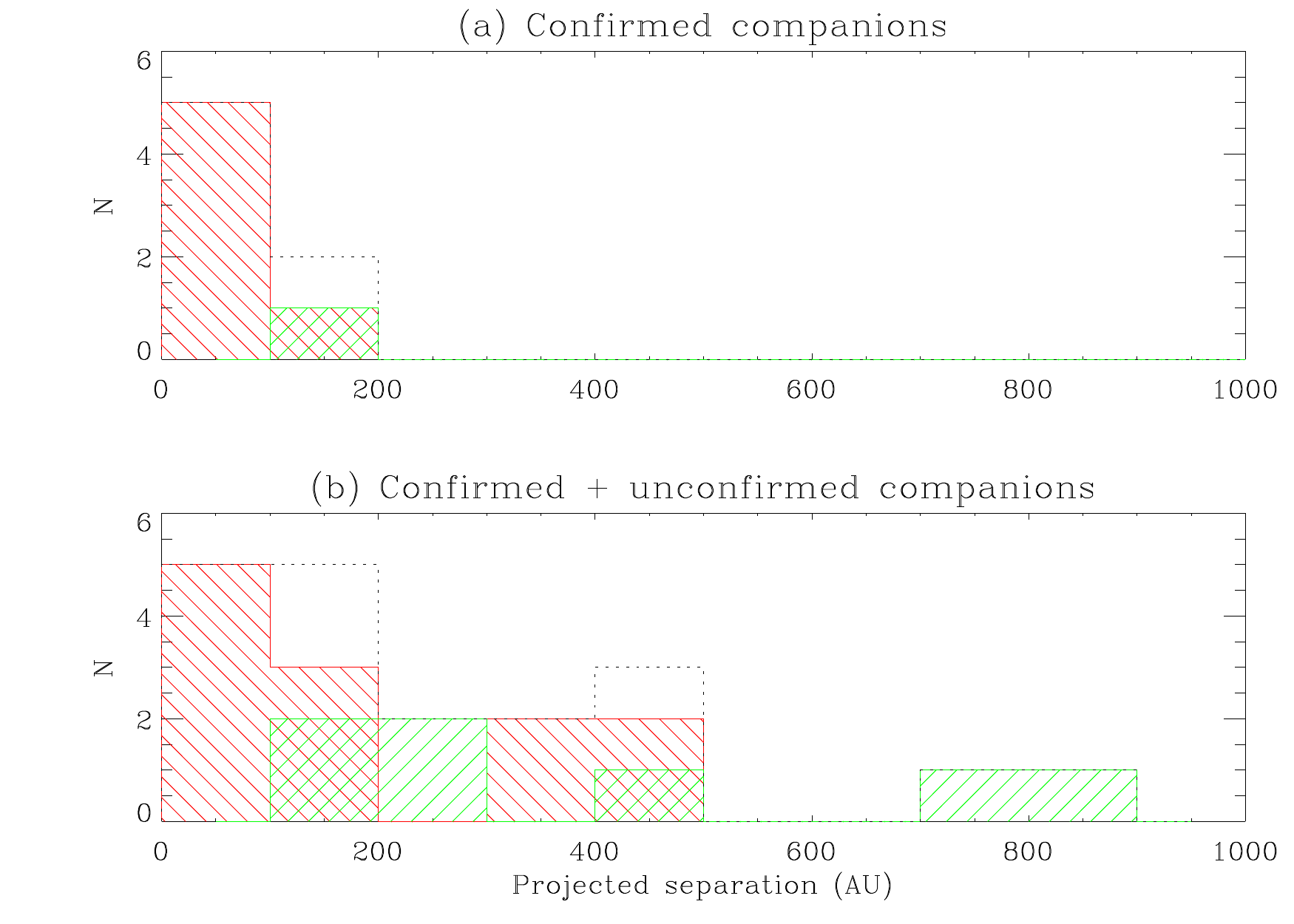}}
\caption{\label{fig:separations} Number of detected companions as a function of the projected separation to the primary, per bin of 100~\au. Companions around F stars (red stripes) and A+B stars (green stripes) are differenciated. The black dotted line represents the total number of companions. (a) Confirmed companions only. (b) Confirmed and unconfirmed companions.}
\end{center}
\end{figure}

Figure~\ref{fig:separations} shows the number of confirmed companions (Fig.~\ref{fig:separations}a) or confirmed \emph{and} unconfirmed companions (Fig.~\ref{fig:separations}b) per bin of 100~\au\ from their primary stars. The bias is clearly apparent in Fig.~\ref{fig:separations}a if we assume that our detection efficiency at $<100$~AU is lower for A stars than for F stars. If we further assume that all unconfirmed companions are physically bound to their primary stars, then Fig.~\ref{fig:separations}b shows that we have roughly the same number of objects between 100 and 500~\au, while companion candidates at projected distances greater than 500~\au\ are only found around A+B stars. It is likely that at least some unconfirmed objects detected around A stars are real companions, and contribute with the observational bias to fill the gap between the multiplicity fractions of A and F stars in the sample.

\subsection{Impact of imaged companions on the velocimetric survey}
The objects from the imaging survey presented in this work have been -- or are being -- monitored spectroscopically, seeking for radial velocity variations that would reveal the presence of close-in companions (see Lagrange \etal\ 2009). Two questions arise when companions are imaged around stars included in the velocimetric survey.

First, is the spectroscopic signal contaminated by the presence of a companion that would appear blended with the primary star in the spectrograph fiber? All background objects detected in the imaging survey are far enough from their stars and do not enter the fiber ($\diameter \sim 2\arcsec$) used in the velocimetric survey. Thus, they do not pollute the radial velocity signal. On the contrary, the companion candidate to \object{HD~43940} (ambiguous case) probably contribute to the velocimetric signal. Six of the seven comoving companion detected, HD~14943B ($\rho = 2\farcs3$), HD~49095BC ($\rho = 1\farcs3$), $\iota$~Crt~B ($\rho = 0\farcs9$), 26~Oph~B ($\rho = 0\farcs3$), and $o$~Gru~B ($\rho = 0\farcs4$), are close enough to their stars to directly contribute to the spectroscopic signal. Note that $o$~Gru~B probably marginally pollutes the velocimetric signal depending on seeing conditions. However, the contribution in terms of flux is generally limited given the flux ratios between the stars and the companions.

On the other hand, are the radial velocity variations a result from the gravitational perturbations of the imaged companions? Indeed, all the 6 stars for which we imaged and confirmed a companion are reported bearing velocimetric variations in Lagrange \etal\ (2009). 
Radial velocity curves (Lagrange \etal\ 2009) indicate that two of them (HD~41742 and $o$~Gru) belong to multiple systems; this includes a spectroscopic binary surrounded by a wider-separation companion, detected only with adaptive optics.\footnote{In the case of HD~41742, there is a 4th companion lying out of the field of view of this survey.} The imaged companions are not responsible in these cases for the radial velocity variations. Meanwhile, the confirmed companions around the other 4 stars partly or totally contribute to the radial velocity variations observed by Lagrange \etal\ (2009) as well as in new unpublished data (Lagrange \etal\, in preparation). In the ambiguous cases of HD~43940 and $\eta$~Hor, the companion candidates could contribute to the observed radial velocity variations, if they are bound objects.

In any case, this shows that the radial velocity curves of at least 6 out of 35 stars followed spectroscopically ($17\%$) may be affected by stellar multiplicity.

\section{Conclusions}
\label{conclu}

We have presented a deep imaging near-infrared survey of 38 austral early-type main sequence stars, searching for low-mass companions at intermediate distances. The overall detection performances obtained for this survey, mainly performed in the K band at 2.2~$\mu$m, allowed us to probe the stellar environments at separations $>0\farcs2$ with contrasts $<0.05$. Farther from the stars ($>5\arcsec$), we achieved limiting contrasts between $10^{-5}$ and $10^{-6}$ using coronagraphy. Although known binary stars were excluded from the survey, we detected 41 companion candidates around 23 stars. Using multi-epoch observations, we were able to determine that 8 candidates are actually comoving companions around 7 stars of the sample, while the other candidates are mainly background contaminants. The comoving companions are low-mass stars with masses ranging from 0.13 to 0.8~\Msun. A comoving binary system consisting of 2 M/L dwarfs imaged around HD~49095, with an almost fully resolved orbit, will be of particular interest for the calibration of low-mass star evolution models. The comoving companions are detected at projected separations ranging from 11.3 to 159~\au. 

At least 16\% of the sample stars, with spectral types ranging from F7 to B9, therefore host a low-mass star companion. This figure is likely a lower limit to the real multiplicity fraction of early-type stars within this separation range. In addition, all the detected and confirmed companions have projected separations below 200~\au: this range matches the physical projected separations of giant planets recently imaged in discs surrounding A-type stars Fomalhaut and HR~8799 (Kalas \etal\ 2008; Marois \etal\ 2008). It is certainly tantalizing to interpret the emerging population of such massive planets as the low-mass tail of the binary distribution of early-type stars. This would imply that such planets are formed by gravitational instability rather than core accretion like closer-in planets. However, further evidence from larger, less-biased surveys, are still required to assess this question in a statistical and quantitative fashion.

Most stars of our the sample are also included in spectroscopic surveys (Lagrange \etal\ 2009) seeking for radial velocity variations. Crossing the results of both surveys show that at least $17\%$ of the radial velocity curves may be affected by companions detected with adaptive optics.

We finally emphasize that tight stellar systems with a bright early-type primary component, such as the ones presented in this study, will be excellent calibrators for next-generation planet imagers such as VLT/SPHERE or interferometers like VLT/Gravity, designed for high-precision astrometry within small field-of-views.

\acknowledgements
We are grateful to the referee J.~Carson, for having promptly provided a useful report on this work. These results have extensively made use of the SIMBAD and VizieR databases, operated at CDS, Strasbourg, France. This publication makes use of data products from the Two Micron All Sky Survey, which is a joint project of the University of Massachusetts and the Infrared Processing and Analysis Center/California Institute of Technology, funded by the National Aeronautics and Space Administration and the National Science Foundation. This research has made use of the Washington Double Star Catalogue maintained at the U.S. Naval Observatory. AML, DE, and GC acknowledge financial support from the French National Research Agency (ANR contract NT05-44463). DE also acknowledges support from the Centre National d'\'Etudes Spatiales (CNES).

\clearpage

\longtab{1}{
\begin{landscape}
\begin{longtable}{rl*{15}{c}}
\hline\hline
\# & \multicolumn{2}{c}{Star name}      & \multicolumn{2}{c}{Proper motions}                 & Spectral & Distance & \multicolumn{4}{c}{Apparent magnitudes} & \multicolumn{2}{c}{Age} & \multicolumn{2}{c}{Multiplicity} & \multicolumn{2}{c}{RV} \\
   &           &                        & $\alpha$       & $\delta$                          & type     &          & $V$   & $J$   & $H$   & $K$            &        & ref.           & code     & ref.                  & code & ref.           \\
   &           &                        & \multicolumn{2}{c}{(mas yr$^{-1}$)}                 &          & (pc)     & (mag) & (mag) & (mag) & (mag)          & (Myr)  &                &          &                       &      &                \\
\hline\endhead
1  & \object{HD   2834} &               & 140.74         & 19.47                    & A0V      & 52.7     & 4.8   & 4.7   & 4.8   & 4.7   & 600--800 &  13            &          &                       & V:SB & 2              \\
2  & \object{HD   3003} & $\beta^3$ Tuc & 86.15          & -49.85                   & A0V      & 46.5     & 5.1   & 5.1   & 5.2   & 5.0   & $\la40$ & 18      & K:B      & 4                     & C    & 2                      \\
3  & \object{HD   4293} &               & -89.06         & -96.11                   & F0V      & 66.6     & 5.9   & 5.4   & 5.3   & 5.2   & 10--100 &  19              &          &                       & C    & 2             \\
4  & \object{HD  10939} &               & 126.45         & 59.04                    & A1V      & 57.0     & 5.0   & 5.0   & 5.0   & 5.0   & 320       & 20               &          &                       &      &              \\
5  & \object{HD  14943} &               & 22.09          & 66.01                    & A5V      & 61.3     & 5.9   & 5.6   & 5.5   & 5.4   & 850    & 1              & N:B      &                       & V    & 2              \\
6  & \object{HD  16555} & $\eta$~Hor    & 111.29         & 1.05                     & A6V      & 44.5     & 5.3   & 4.6   & 4.6   & 4.5   & 50--350 & 21             &          &                       & V    & 3              \\
7  & \object{HD  16754} & s Eri         & 92.11          & -15.07                   & A1V      & 39.8     & 4.8   &          &         &          & $\sim350$ & 22  & K:B      & 4                     &      &                \\
8  & \object{HD  19545} &               & 65.72          & -20.57                   & A5V      & 57.6     & 6.2   & 5.9   & 5.9   & 5.8   & $\sim27$   & 23  &          &                       & V:SB & 2              \\
9  & \object{HD  20888} &               & 57.15          & 13.07                    & A3V      & 58.0     & 6.0   & 5.8   & 5.8   & 5.7   & 10--40 & 5              &          &                       &      &                \\
10 & \object{HD  21882} &               & -59.48         & -1.45                    & A5V      & 62.3     & 5.8   & 5.4   & 5.2   & 5.2   &        &                &          &                       & C    & 2              \\
11 & \object{HD  29875} & $\alpha$ Cae  & -141.18        & -74.95                   & F2V      & 20.1     & 4.5   & 3.9   & 3.7   & 3.7   & 1\,000 & 6              & K:B      & 4                     & V    & 2              \\
12 & \object{HD  29992} &               & 46.9           & 193.14                   & F3V      & 27.7     & 5.1   & 4.7   & 4.3   & 4.1   &  1\,600& 6               &          &                       & V    & 2              \\
13 & \object{HD  31746} &               & 100.99         & 68.25                    & F5V      & 30.9     & 6.1   & 5.3   & 5.1   & 5.0   & 1\,400 & 6              &          &                       & V    & 2              \\
14 & \object{HD  32743} & $\eta^1$ Pic  & -43.96         & 27.18                    & F5V      & 26.2     & 5.4   & 4.8   & 4.5   & 4.4   & 1\,400 & 6              & K:B      & 4                     & V    & 2              \\
15 & \object{HD  41742} &               & -79.98         & 254.28                   & F4V      & 27.0     & 5.9   & 5.1   & 4.9   & 4.7   & 3\,700 & 6              & K:T      & 7                     & V:SB & 2              \\
16 & \object{HD  43940} &               & -5.51          & -22.63                   & A2V      & 62.1     & 5.9   & 5.6   & 5.5   & 5.4   &        &                & N:B      &                       & C    & 2              \\
17 & \object{HD  49095} &               & -204.91        & -304.12                  & F6.5V    & 24.3     & 5.9   & 5.0   & 4.8   & 4.7   & 2\,000 & 6              & K:B,N:T  & 8                     & V    & 2              \\
18 & \object{HD  50445} &               & -40.47         & -59.14                   & A3V      & 55.1     & 5.9   & 5.6   & 5.5   & 5.5   &        &                & N:M      &                       & C    & 2              \\
19 & \object{HD  68456} &               & -155.04        & -297.68                  & F6V      & 21.4     & 4.8   & 3.9   & 3.6   & 3.7   & 2\,500 & 6              & K:B      & 12                    & V:SB & 2              \\
20 & \object{HD  75171} &               & -63.56         & 104.15                   & A4V      & 60.6     & 6.0   & 5.6   & 5.5   & 5.5   & 800    & 13             &          &                       & V    & 2              \\
21 & \object{HD  91889} & 	           & 268.49	        & -672.31                  & F7V      & 24.6     & 5.7   & 4.9   & 4.5   & 4.3   & 4\,100 & 6              &          &                       & V    & 2              \\
22 & \object{HD 101198} & $\iota$ Crt   & 99.62          & 126.17                   & F6.5V    & 27.0     & 5.5   & 4.8   & 4.3   & 4.1   & 4\,800 & 6              & K:B      & 4                     & V    & 2              \\
23 & \object{HD 112934} &               & -118.89        & -59.5                    & A9V      & 54.9     & 6.6   & 6.0   & 5.9   & 5.8   & 2\,000 & 6              &          &                       & V:SB & 2              \\
24 & \object{HD 116568} & 66 Vir        & 157.73         & -36.24                   & F3.5V    & 30.1     & 5.8   & 5.3   & 4.9   & 4.7   & 2\,000 & 6              &          &                       & V:SB & 2              \\
25 & \object{HD 153363} & 26 Oph        & 49.14          & -54.89                   & F3V      & 33.3     & 5.7   & 5.0   & 4.8   & 4.7   & 1\,700 & 6              & K:B      & 8,14                  & V    & 2              \\
26 & \object{HD 158094} & $\delta$ Ara  & -53.65         & -99.37                   & B8V      & 57.4     & 3.6   & 3.7   & 3.7   & 3.7   & 125    & 15             & K:B      & 4                     & V:B  & 2              \\
27 & \object{HD 177756} & $\lambda$ Aql & -19.68         & -90.37                   & B9V      & 38.4     & 3.4   & 3.5   & 3.5   & 3.6   & 90     & 24             &          &                       & V:B  & 2              \\
28 & \object{HD 186543} & $\nu$ Tel     & 92.4           & -137.4                   & A7III-IV & 52.1     & 5.3   & 5.0   & 5.0   & 4.9   & 250    & 25             &          &                       & V    & 2              \\
29 & \object{HD 197692} & $\psi$ Cap    & -51.38         & -156.66                  & F5V      & 14.7     & 4.2   & 3.4   & 3.1   & 3.1   & 1\,400 & 6              &          &                       & V    & 2              \\
30 & \object{HD 200761} & $\theta$ Cap  & 79.64          & -61.64                   & A1V      & 48.5     & 4.1   & 4.4   & 4.3   & 4.1   &        &                &          &                       & V:B  & 2              \\
31 & \object{HD 209819} & $\iota$ Aqr   & 40.45          & -57.16                   & B8V      & 52.9     & 4.3   & 4.4   & 4.6   & 4.4   & 30--60 & 26             &          &                       & V:B  & 2              \\
32 & \object{HD 213398} & $\beta$ PsA   & 59.64          & -18.7                    & A0V      & 45.5     & 4.3   & 4.3   & 4.3   & 4.3   & 240    & 24             & K:B      & 4                     & C    & 2              \\
33 & \object{HD 216385} & $\sigma$ Peg A& 521.86         & 44.1                     & F7IV     & 26.8     & 5.2   & 4.2   & 3.9   & 3.9   & 2\,700 & 6              & K:B      & 16                    & V    & 9              \\
34 & \object{HD 216627} & $\delta$ Aqr  & -44.08         & -24.81                   & A3V      & 48.9     & 3.3   & 3.3   & 3.2   & 3.2   & 300    & 13,17          & K:B      & 12                    & V:SB & 2              \\
35 & \object{HD 219482} &               & 176            & -26.12                   & F6V      & 20.6     & 5.7   & 5.1   & 4.6   & 4.4   & 3\,700 & 6              &          &                       & V    & 2              \\
36 & \object{HD 220729} & $o$ Gru       & 33.16          & 130.07                   & F4V      & 32.0     & 5.5   & 4.9   & 4.7   & 4.5   & 1\,100 & 6              & N:B      &                       & V:SB & 2              \\
37 & \object{HD 223011} &               & 103.11         & -36.01                   & A7III-IV & 65.3     & 6.3   & 5.9   & 5.9   & 5.8   &        &                &          &                       & V    & 2              \\
38 & \object{HD 224392} & $\eta$ Tuc    & 78.86          & -61.1                    & A1V      & 48.7     & 5.0   & 4.9   & 4.9   & 4.8   & 10--40 & 18             &          &                       & C    & 2              \\
\hline
\caption{\label{tab:sample}
Star sample for the southern survey. Infrared $J$, $H$, and $K$ magnitudes are extracted from 2MASS (Skrutskie \etal\ 2006). The `RV' column indicates the radial velocimetry status (C: constant, V: variable, with possible source of variations being SB: spectral binary, A: magnetic activity, Pu: pulsations, Pl: planets, D: drift) as determined by Lagrange \etal\ 2009 with HARPS or as relying on SOPHIE data (Desort \etal, personal communication). The column `Multiplicity' indicates apparent (or comoving) binary (B), ternary (T), or more-component (M) systems, whether it is new (N) or known before this survey (K). References are given below the table.\newline
References: (1) Su \etal\ 2006; (2) HARPS survey, Lagrange \etal\ 2009; (3) HARPS survey, unpublished data; (4) CCDM II, Vizier Online Data Catalogue I/274, Dommanget \& Nys 2002; (5) Fern\'andez, Figueras \& Torra 2008; (6) Geneva-Copenhagen Survey of Solar neighbourhood III, Vizier Online Data Catalogue V/130, Holmberg, Nordstr\"om \& Andersen 2009; (7) Multiple Star Catalog, Tokovinin 1997; (8) Proper motion derivatives of binaries, Makarov \& Kaplan 2005; (9) SOPHIE survey, unpublished data; (10) $\rm S_{B^9}$, Pourbaix \etal\ 2004; (11) Desort \etal\ 2008; (12) Goldin \& Makarov 2007; (13) Eggen 1998; (14) proper motion binary, Frankowski \etal\ 2007; (15) Rieke \etal\ 2005; (16) The Washington Visual Double Star Catalogue, Mason \etal\ 2001; (17) King \etal\ 2003; (18) member of the Tucana association, Zuckermann, Song \& Webb 2001; (19) HD~4293 is classified as a F0V but has colours corresponding to a F0III giant (Hauck 1986) with typical age between 10 and 100~Myr (Eggen 1991a); (20) Morales \etal\ 2009; (21) Plavchan \etal\ 2009; (22) Gray \etal\ 2006; (23) Makarov 2007 suggested that HD~19545 and \object{BO~Microscopii} (`Speedy Mic') were formerly part of the same system within the Tucana-Horologium association; (24) Grosb\o l 1978; (25) member of the IC~2391 supercluster (Eggen 1991b); (26) a field B star, Huang \& Gies 2008}
\end{longtable}
\end{landscape}
}
\clearpage

\longtab{3}{
\begin{longtable}{rl*{7}{c}}
\hline\hline
\#& Star name   & Epoch & Observing date & Telescope/Instrument & Camera  & Filter & Mode & CC \# \\
\hline\endhead
1 & HD   2834   & II    & 2005-11-06     & VLT/NaCo             & S27     & \Ks    & D+C  & 0 \\
2 & HD   3003   & II    & 2005-11-07     & VLT/NaCo             & S27     & \Ks    & D+C  & 0 \\
3 & HD   4293   & II    & 2005-11-08     & VLT/NaCo             & S27     & \Ks    & D+C  & 0 \\
4 & HD  10939   & II    & 2005-11-06     & VLT/NaCo             & S27     & \Ks    & D+C  & 0 \\
5 & HD  14943   & II    & 2005-11-07     & VLT/NaCo             & S27     & \Ks    & D+C  & 1 \\
  &             & V     & 2008-08-20     & VLT/NaCo             & S13,S27 & \Ks    & D    & 1 \\
6 & HD  16555   & II    & 2005-11-06     & VLT/NaCo             & S27     & \J,\Ks & D+C  & 1 \\
  &             & V     & 2008-08-20     & VLT/NaCo             & S27     & \Ks    & D+C  & 1 \\
7 & HD  16754   & II    & 2005-11-07     & VLT/NaCo             & S27     & \Ks    & D+C  & 0 \\
8 & HD  19545** & V     & 2008-08-20     & VLT/NaCo             & S27     & \Ks    & D    & 0 \\
9 & HD  20888*  & II    & 2005-11-07     & VLT/NaCo             & S27     & \Ks    & D+C  & 1 \\
10& HD  21882   & II    & 2005-11-06     & VLT/NaCo             & S13,S27 & \J,\Ks & D+C  & 1 \\
  &             & V     & 2008-08-21     & VLT/NaCo             & S27     & \Ks    & D+C  & 1 \\
11& HD  29875*  & II    & 2005-11-08     & VLT/NaCo             & S27     & \Ks    & D+C  & 2 \\
12& HD  29992   & V     & 2008-08-20     & VLT/NaCo             & S27     & \Ks    & D+C  & 0 \\
13& HD  31746   & II    & 2005-11-07     & VLT/NaCo             & S27     & \Ks    & D+C  & 0 \\
14& HD  32743*  & II    & 2005-11-07     & VLT/NaCo             & S27     & \J,\Ks & D+C  & 4 \\
15& HD  41742   & II    & 2005-11-06     & VLT/NaCo             & S27     & \Ks    & D+C  & 3 \\
  &             & IV    & 2007-11-17     & CFHT/PUEO            & KIR     & \BrG   & D+S  & 3 \\
  &             & V     & 2008-08-21     & VLT/NaCo             & S27     & \Ks    & D+C  & 3 \\
  &             & VI    & 2009-04-25     & VLT/NaCo             & S27     & \Ks    & D    & 1 \\
16& HD  43940   & II    & 2005-11-08     & VLT/NaCo             & S27     & \Ks    & D+C  & 1 \\
  &             & V     & 2008-08-20     & VLT/NaCo             & S13,S27 & \H,\Ks & D    & 1?\\
  &             & VI    & 2009-04-26     & VLT/NaCo             & S13     & \J     & D    & 0 \\
17& HD  49095   & II    & 2005-11-07     & VLT/NaCo             & S27     & \Ks    & D+C  & 3 \\
  &             & III   & 2007-01-27     & CFHT/PUEO            & KIR     & \Kp,\BrG & D+S& 2 \\
  &             & IV    & 2007-11-16     & CFHT/PUEO            & KIR     & \FeII  & D    & 2 \\
  &             & VI    & 2009-04-26     & VLT/NaCo             & S13     & \H     & D    & 2 \\
18& HD  50445   & II    & 2005-11-06     & VLT/NaCo             & S27     & \Ks    & D+C  & 5 \\
  &             & VI    & 2009-04-26     & VLT/NaCo             & S27     & \Ks    & C*** & 5 \\
19& HD  68456   & II    & 2005-11-07     & VLT/NaCo             & S27     & \Ks    & D+C  & 1 \\
  &             & VI    & 2009-04-26     & VLT/NaCo             & S27     & \Ks    & D+C  & 1 \\
20& HD  75171*  & II    & 2005-11-08     & VLT/NaCo             & S27     & \Ks    & D+C  & 2 \\
21& HD  91889   & I     & 2005-01-27     & CFHT/PUEO            & KIR     & \BrG   & D+S  & 1 \\
  &             & VI    & 2009-04-26     & VLT/NaCo             & S27     & \Ks    & D+C  & 1 \\
22& HD 101198   & IV    & 2007-11-16     & CFHT/PUEO            & KIR     & \FeII  & D    & 1 \\
  &             & VI    & 2009-04-26     & VLT/NaCo             & S13,S27 & \H,\Ks & D    & 1 \\
23& HD 112934** & III   & 2007-01-28     & CFHT/PUEO            & KIR     & K      & D+S  & 0?\\
  &             & V     & 2008-08-20     & VLT/NaCo             & S27     & \Ks    & D    & 0 \\
24& HD 116568** & V     & 2008-08-20     & VLT/NaCo             & S27     & \Ks    & D    & 0 \\
25& HD 153363   & V     & 2008-08-20     & VLT/NaCo             & S27     & \Ks    & D+C  & M\\
  &             & VI    & 2009-04-26     & VLT/NaCo             & S13     & \J,\H  & D    & 2\\
26& HD 158094   & V     & 2008-08-21     & VLT/NaCo             & S27     & \Ks    & D    & 0\\
  &             & VII   & 2009-08-27     & VLT/NaCo             & S27     & \Ks    & D+C  & 3\\
27& HD 177756** & V     & 2008-08-21     & VLT/NaCo             & S27     & \Ks    & D    & 1\\
  &             & VI    & 2009-04-26     & VLT/NaCo             & S13,S27 & \H,\NBdeuxdixsept & D    & 1\\
28& HD 186543   & V     & 2008-08-20     & VLT/NaCo             & S27     & \Ks    & D+C  & 0\\
29& HD 197692   & II    & 2005-11-07     & VLT/NaCo             & S27     & \Ks    & D+C  & 1\\
  &             & V     & 2008-08-21     & VLT/NaCo             & S27     & \Ks    & D+C  & 1\\
30& HD 200761   & V     & 2008-08-21     & VLT/NaCo             & S27     & \Ks    & D    & 0\\
  &             & VII   & 2009-08-27     & VLT/NaCo             & S27     & \Ks    & D+C  & 0\\
31& HD 209819   & V     & 2008-08-21     & VLT/NaCo             & S27     & \Ks    & D    & 0\\
  &             & VII   & 2009-08-27     & VLT/NaCo             & S27     & \Ks    & D+C  & 0\\
32& HD 213398   & II    & 2005-11-06     & VLT/NaCo             & S27     & \J,\Ks & D+C  & 1\\
  &             & V     & 2008-08-20     & VLT/NaCo             & S27     & \Ks    & D+C  & 1\\
33& HD 216385   & IV    & 2007-11-16     & CFHT/PUEO            & KIR     & \BrG,\FeII & D+S & 1\\
  &             & V     & 2008-08-21     & VLT/NaCo             & S27     & \Ks    & D+C  & 1\\
34& HD 216627   & V     & 2008-08-21     & VLT/NaCo             & S27     & \Ks    & D    & 0\\
  &             & VII   & 2009-08-27     & VLT/NaCo             & S27     & \NBdeuxdouze   & D+C  & 0\\
35& HD 219482   & V     & 2008-08-21     & VLT/NaCo             & S27     & \Ks    & D+C  & 1\\
  &             & VI    & 2009-04-26     & VLT/NaCo             & S27     & \Ks    & D+C  & 1\\
36& HD 220729   & II    & 2005-11-07     & VLT/NaCo             & S27     & \Ks    & D+C  & 1\\
  &             & V     & 2008-08-20     & VLT/NaCo             & S13,S27 & \H,\Ks & D    & 1\\
37& HD 223011** & V     & 2008-08-20     & VLT/NaCo             & S27     & \Ks    & D    & 0?\\    
38& HD 224392   & II    & 2005-11-07     & VLT/NaCo             & S27     & \Ks    & D+C  & 2\\
  &             & V     & 2008-08-21     & VLT/NaCo             & S27     & \Ks    & D+C  & 2\\
\hline
\caption{\label{tab:obs}
Observing setups, dates, and number of companion candidates observed (`M' stands for many). For each observing date, we precise the observing mode (D: direct imaging, C: coronagraphic imaging, S: saturated direct imaging), the filter(s) (NaCo filters: \J, \H, \Ks, \NBdeuxdouze; PUEO filters: K', \ion{Fe}{ii}, Br$\gamma$), and camera(s) used (NaCo cameras: S27, S13; PUEO camera: KIR).  The last column `Companionship' indicates the presence of companion candidate(s) (CC) or bound companion(s) (BC) when this has been established at several epochs. The epochs II, V, VI, and VII correspond to ESO observing programmes 076.C-0270(A), 081.C-0653(A), 083.C-0151(B), and 083.C-0151(A), respectively. Epochs I, III, and IV correspond to CFHT programmes \ldots  \newline
*No second-epoch observation was recorded.\newline
**No coronagraphic image was recorded.\newline
}
\end{longtable}
}
\clearpage


\longtab{5}{
\begin{landscape}
\begin{longtable}{rl*{10}{c}}
\hline \hline
\#& Star     & CC\# & Date       & Filter & Camera & Mode & $\rm DIT\times NDIT\times NEXP$ & Projected separation & Position angle   & $\Delta m\pm0.4$ & Status \\
  &          &      &            &        &        &      & (s)                             & $\rho$ (\arcsec)     & $\theta$ (\degr) & (mag)            &        \\
\hline\endhead
5 & HD014943 & 1 & 2005-11-07 & Ks & S27 & D & $1.0\times5\times10$ & $2.2921\pm0.0107$ & $153.970\pm0.54$ & $5.10$ & Comoving \\
5 & HD014943 & 1 & 2008-08-20 & Ks & S27 & D & $2.0\times7\times10$ & $2.2830\pm0.0104$ & $154.080\pm0.53$ & $5.10$ & Comoving \\
5 & HD014943 & 1 & 2008-08-20 & Ks & S13 & D & $5.0\times3\times10$ & $2.2843\pm0.0075$ & $153.840\pm0.41$ & $5.20$ & Comoving \\[4pt]
6 & HD016555 & 1 & 2005-11-06 & Ks & S27 & C & $2.0\times15\times20$ & $4.8718\pm0.0118$ & $-130.14\pm0.77$ & $9.00$ & Ambiguous \\
6 & HD016555 & 1 & 2008-08-20 & Ks & S27 & C & $2.0\times15\times10$ & $4.9323\pm0.0107$ & $-127.67\pm0.44$ & $10.0$ & Ambiguous \\
6 & HD016555 & 1 & 2008-08-20 & Ks & S27 & C & $2.0\times15\times10$ & $4.9279\pm0.0107$ & $-127.55\pm0.44$ & $10.0$ & Ambiguous \\[4pt]
9 & HD020888 & 1 & 2005-11-07 & Ks & S27 & D & $0.8\times5\times10$ & $4.9044\pm0.0136$ & $17.7900\pm0.19$ & $3.80$ & Undefined \\[4pt]
10 & HD021882 & 1 & 2005-11-06 & Ks & S27 & C & $6.0\times5\times18$ & $5.3400\pm0.0128$ & $-60.270\pm0.18$ & $11.3$ & Background \\
10 & HD021882 & 1 & 2008-08-21 & Ks & S27 & C & $8.0\times4\times10$ & $5.1903\pm0.0110$ & $-58.970\pm0.17$ & $10.4$ & Background \\[4pt]
11 & HD029875 & 1 & 2005-11-08 & Ks & S27 & D & $0.3\times5\times10$ & $7.0950\pm0.0140$ & $140.820\pm0.61$ & $3.90$ & Undefined \\[2pt]
11 & HD029875 & 2 & 2005-11-08 & Ks & S27 & D & $0.3\times5\times10$ & $7.4930\pm0.0143$ & $137.260\pm1.43$ & $4.20$ & Undefined \\[4pt]
14 & HD032743 & 1 & 2005-11-07 & Ks & S27 & D & $0.4\times5\times10$ & $12.212\pm0.0285$ & $-176.32\pm0.12$ & $7.40$ & Undefined \\[2pt]
14 & HD032743 & 2 & 2005-11-07 & Ks & S27 & C & $6.0\times5\times18$ & $12.268\pm0.0291$ & $-177.38\pm0.12$ & $9.10$ & Undefined \\[2pt]
14 & HD032743 & 3 & 2005-11-07 & Ks & S27 & C & $6.0\times5\times18$ & $15.639\pm0.0243$ & $-39.430\pm0.12$ & $11.7$ & Undefined \\[2pt]
14 & HD032743 & 4 & 2005-11-07 & Ks & S27 & C & $6.0\times5\times18$ & $15.592\pm0.0242$ & $-39.760\pm0.12$ & $12.2$ & Undefined \\[4pt]
15 & HD041742 & 1 & 2005-11-06 & Ks & S27 & D & $0.4\times5\times10$ & $5.8393\pm0.0129$ & $-144.35\pm0.45$ & $1.80$ & Comoving \\
15 & HD041742 & 1 & 2007-11-17 & BrG & KIR & D & $1.5\times1\times25$ & $5.8668\pm0.0130$ & $-147.24\pm0.37$ & $1.70$ & Comoving \\
15 & HD041742 & 1 & 2008-08-21 & Ks & S27 & D & $0.4\times25\times10$ & $5.8710\pm0.0112$ & $-144.46\pm0.40$ & $1.70$ & Comoving \\
15 & HD041742 & 1 & 2009-04-25 & Ks & S27 & D & $0.4\times30\times14$ & $5.8780\pm0.0104$ & $-145.06\pm0.35$ & $1.50$ & Comoving \\[2pt]
15 & HD041742 & 2 & 2005-11-06 & Ks & S27 & C & $4.0\times7\times3$ & $7.5084\pm0.0188$ & $98.4100\pm0.14$ & $8.10$ & Background \\
15 & HD041742 & 2 & 2007-11-17 & BrG & KIR & S & $40.\times1\times25$ & $7.7782\pm0.0164$ & $98.9100\pm0.13$ & $10.6$ & Background \\
15 & HD041742 & 2 & 2008-08-21 & Ks & S27 & C & $4.0\times8\times10$ & $7.8713\pm0.0136$ & $103.140\pm0.13$ & $7.50$ & Background \\[2pt]
15 & HD041742 & 3 & 2005-11-06 & Ks & S27 & C & $4.0\times7\times3$ & $8.4293\pm0.0207$ & $-6.8000\pm0.14$ & $8.80$ & Background \\
15 & HD041742 & 3 & 2007-11-17 & BrG & KIR & S & $40.\times1\times25$ & $7.8699\pm0.0165$ & $-8.5500\pm0.13$ & $10.9$ & Background \\
15 & HD041742 & 3 & 2008-08-21 & Ks & S27 & C & $4.0\times8\times10$ & $7.6914\pm0.0147$ & $-5.5000\pm0.13$ & $7.90$ & Background \\[4pt]
16 & HD043940 & 1 & 2005-11-08 & Ks & S27 & D & $1.0\times5\times10$ & $0.2186\pm0.0099$ & $-148.65\pm6.86$ & $2.10$ & Ambiguous \\
16 & HD043940 & 1 & 2008-08-20 & Ks & S27 & D & $2.0\times5\times10$ & $0.0441\pm0.0096$ & $-132.47\pm135.$ & $0.10$ & Ambiguous \\
16 & HD043940 & 1 & 2008-08-20 & H & S13 & D & $5.0\times2\times10$ & $0.0469\pm0.0050$ & $-154.89\pm11.9$ & $0.40$ & Ambiguous \\[4pt]
17 & HD049095 & 1 & 2005-11-07 & Ks & S27 & D & $0.4\times5\times10$ & $1.3173\pm0.0099$ & $-145.44\pm1.42$ & $5.40$ & Comoving \\
17 & HD049095 & 1 & 2007-01-27 & BrG & KIR & D & $3.0\times1\times35$ & $1.3294\pm0.0127$ & $-147.61\pm1.55$ & $4.80$ & Comoving \\
17 & HD049095 & 1 & 2007-11-16 & FeII & KIR & D & $1.0\times1\times25$ & $1.2825\pm0.0127$ & $-149.12\pm1.47$ & $4.70$ & Comoving \\
17 & HD049095 & 1 & 2009-04-26 & H & S13 & D & $0.8\times20\times21$ & $1.3605\pm0.0071$ & $-145.99\pm1.00$ & $5.60$ & Comoving \\[2pt]
17 & HD049095 & 2 & 2005-11-07 & Ks & S27 & D & $0.4\times5\times10$ & $1.2890\pm0.0098$ & $-140.94\pm2.34$ & $5.60$ & Comoving \\
17 & HD049095 & 2 & 2007-01-27 & BrG & KIR & D & $3.0\times1\times35$ & $1.2526\pm0.0128$ & $-148.98\pm1.53$ & $4.90$ & Comoving \\
17 & HD049095 & 2 & 2007-11-16 & FeII & KIR & D & $1.0\times1\times25$ & $1.3404\pm0.0127$ & $-149.04\pm1.41$ & $4.90$ & Comoving \\
17 & HD049095 & 2 & 2009-04-26 & H & S13 & D & $0.8\times20\times21$ & $1.3284\pm0.0071$ & $-149.64\pm0.82$ & $6.10$ & Comoving \\[4pt]
18 & HD050445 & 1 & 2005-11-06 & Ks & S27 & C & $5.0\times6\times20$ & $12.698\pm0.0268$ & $80.2700\pm0.12$ & $8.30$ & Background \\
18 & HD050445 & 1 & 2009-04-26 & Ks & S27 & C & $5.0\times4\times10$ & $12.903\pm0.0143$ & $79.1700\pm0.08$ & $8.50$ & Background \\[2pt]
18 & HD050445 & 2 & 2005-11-06 & Ks & S27 & C & $5.0\times6\times20$ & $15.073\pm0.0256$ & $154.050\pm0.18$ & $8.00$ & Background \\
18 & HD050445 & 2 & 2009-04-26 & Ks & S27 & C & $5.0\times4\times10$ & $14.952\pm0.0138$ & $152.750\pm0.12$ & $7.90$ & Background \\[2pt]
18 & HD050445 & 3 & 2005-11-06 & Ks & S27 & C & $5.0\times6\times20$ & $14.221\pm0.0252$ & $157.420\pm0.16$ & $10.6$ & Background \\
18 & HD050445 & 3 & 2009-04-26 & Ks & S27 & C & $5.0\times4\times10$ & $14.094\pm0.0136$ & $156.090\pm0.11$ & $10.4$ & Background \\[2pt]
18 & HD050445 & 4 & 2005-11-06 & Ks & S27 & C & $5.0\times6\times20$ & $7.9742\pm0.0187$ & $101.490\pm0.14$ & $11.7$ & Background \\
18 & HD050445 & 4 & 2009-04-26 & Ks & S27 & C & $5.0\times4\times10$ & $8.1163\pm0.0129$ & $99.5500\pm0.11$ & $11.8$ & Background \\[2pt]
18 & HD050445 & 5 & 2005-11-06 & Ks & S27 & C & $5.0\times6\times20$ & $10.474\pm0.0205$ & $-109.21\pm0.13$ & $12.7$ & Background \\
18 & HD050445 & 5 & 2009-04-26 & Ks & S27 & C & $5.0\times4\times10$ & $10.304\pm0.0125$ & $-108.88\pm0.10$ & $12.3$ & Background \\[4pt]
19 & HD068456 & 1 & 2005-11-07 & Ks & S27 & C & $1.6\times15\times20$ & $5.4116\pm0.0134$ & $66.3800\pm0.18$ & $10.3$ & Background \\
19 & HD068456 & 1 & 2009-04-26 & Ks & S27 & C & $1.0\times3\times20$ & $6.4428\pm0.0106$ & $58.2400\pm0.13$ & $10.9$ & Background \\[4pt]
20 & HD075171 & 1 & 2005-11-08 & Ks & S27 & C & $10.\times3\times20$ & $4.5346\pm0.0123$ & $-26.070\pm0.20$ & $10.3$ & Undefined \\[2pt]
20 & HD075171 & 2 & 2005-11-08 & Ks & S27 & C & $10.\times3\times20$ & $14.208\pm0.0285$ & $-167.82\pm0.13$ & $11.1$ & Undefined \\[4pt]
21 & HD091889 & 1 & 2005-01-27 & BrG & KIR & S & $40.\times1\times15$ & $9.4738\pm0.0225$ & $-64.600\pm0.16$ & $12.3$ & Background \\
21 & HD091889 & 1 & 2009-04-26 & Ks & S27 & C & $2.0\times3\times10$ & $11.987\pm0.0123$ & $-53.050\pm0.08$ & $10.0$ & Background \\[4pt]
22 & HD101198 & 1 & 2007-11-16 & FeII & KIR & D & $0.5\times1\times25$ & $0.8912\pm0.0136$ & $-109.64\pm1.16$ & $3.10$ & Comoving \\
22 & HD101198 & 1 & 2009-04-26 & H & S13 & D & $0.5\times20\times12$ & $0.9264\pm0.0063$ & $-108.77\pm0.55$ & $3.10$ & Comoving \\
22 & HD101198 & 1 & 2009-04-26 & Ks & S27 & D & $0.3\times20\times12$ & $0.9270\pm0.0106$ & $-109.07\pm0.86$ & $3.00$ & Comoving \\[4pt]
25 & HD153363 & 1 & 2008-08-20 & Ks & S27 & D & $2.0\times5\times10$ & $0.3341\pm0.0115$ & $-78.740\pm2.32$ & $2.00$ & Comoving \\
25 & HD153363 & 1 & 2009-04-26 & H & S13 & D & $0.8\times5\times60$ & $0.3598\pm0.0061$ & $-83.470\pm1.11$ & $2.70$ & Comoving \\
25 & HD153363 & 1 & 2009-04-26 & J & S13 & D & $0.8\times5\times20$ & $0.3596\pm0.0061$ & $-83.680\pm1.08$ & $3.10$ & Comoving \\[2pt]
25 & HD153363 & 2 & 2008-08-20 & Ks & S27 & D & $2.0\times5\times10$ & $8.8886\pm0.0131$ & $-117.21\pm0.14$ & $8.00$ & Background \\
25 & HD153363 & 2 & 2009-04-26 & H & S13 & D & $0.8\times5\times60$ & $8.6715\pm0.0342$ & $-117.07\pm0.39$ & $7.60$ & Background \\[4pt]
26 & HD158094 & 1 & 2009-08-27 & Ks & S27 & C & $0.5\times60\times10$ & $3.1229\pm0.0105$ & $67.2100\pm0.25$ & $10.5$ & Undefined \\[2pt]
26 & HD158094 & 2 & 2009-08-27 & Ks & S27 & C & $0.5\times60\times10$ & $8.5888\pm0.0108$ & $51.0100\pm0.10$ & $9.30$ & Undefined \\[2pt]
26 & HD158094 & 3 & 2009-08-27 & Ks & S27 & C & $0.5\times60\times10$ & $12.317\pm0.0123$ & $62.0300\pm0.08$ & $8.20$ & Undefined \\[4pt]
27 & HD177756 & 1 & 2008-08-21 & Ks & S27 & D & $1.0\times10\times10$ & $12.448\pm0.0154$ & $30.5400\pm0.10$ & $4.00$ & Background \\
27 & HD177756 & 1 & 2009-04-26 & Ks & S27 & D & $1.5\times20\times8$ & $12.459\pm0.0127$ & $29.7900\pm0.08$ & $3.80$ & Background \\[4pt]
29 & HD197692 & 1 & 2005-11-07 & Ks & S27 & C & $0.8\times30\times20$ & $11.040\pm0.0194$ & $60.8100\pm0.13$ & $12.6$ & Background \\
29 & HD197692 & 1 & 2008-08-21 & Ks & S27 & C & $0.5\times80\times10$ & $11.418\pm0.0147$ & $59.4600\pm0.11$ & $12.2$ & Background \\[4pt]
32 & HD213398 & 1 & 2005-11-06 & Ks & S27 & C & $3.0\times9\times20$ & $9.3478\pm0.0163$ & $-43.620\pm0.14$ & $10.8$ & Background \\
32 & HD213398 & 1 & 2005-11-06 & J & S27 & C & $5.0\times6\times20$ & $9.3574\pm0.0163$ & $-43.630\pm0.14$ & $9.50$ & Background \\
32 & HD213398 & 1 & 2008-08-20 & Ks & S27 & C & $2.0\times15\times10$ & $9.5217\pm0.0132$ & $-44.080\pm0.11$ & $10.7$ & Background \\[4pt]
33 & HD216385 & 1 & 2007-11-16 & BrG & KIR & S & $0.0\times1\times30$ & $15.985\pm0.0199$ & $83.1000\pm0.07$ & $6.30$ & Background \\[2pt]
33 & HD216385 & 2 & 2008-08-21 & Ks & S27 & C & $1.5\times20\times10$ & $14.685\pm0.0171$ & $125.720\pm0.18$ & $12.4$ & Background \\[4pt]
35 & HD219482 & 1 & 2008-08-21 & Ks & S27 & C & $4.0\times8\times10$ & $10.298\pm0.0154$ & $-10.240\pm0.12$ & $11.7$ & Background \\
35 & HD219482 & 1 & 2009-04-26 & Ks & S27 & C & $2.5\times6\times10$ & $10.192\pm0.0132$ & $-11.550\pm0.10$ & $12.3$ & Background \\[4pt]
36 & HD220729 & 1 & 2005-11-07 & Ks & S27 & D & $0.8\times5\times10$ & $0.4757\pm0.0102$ & $115.270\pm1.84$ & $4.70$ & Comoving \\
36 & HD220729 & 1 & 2008-08-20 & Ks & S27 & D & $1.0\times10\times10$ & $0.4435\pm0.0100$ & $118.480\pm2.17$ & $4.60$ & Comoving \\
36 & HD220729 & 1 & 2008-08-20 & H & S13 & D & $4.0\times3\times10$ & $0.4548\pm0.0049$ & $121.650\pm1.24$ & $4.40$ & Comoving \\[4pt]
38 & HD224392 & 1 & 2005-11-07 & Ks & S27 & C & $0.0\times3\times20$ & $13.249\pm0.0251$ & $107.390\pm0.12$ & $7.50$ & Background \\
38 & HD224392 & 1 & 2008-08-21 & Ks & S27 & C & $0.0\times3\times20$ & $13.041\pm0.0165$ & $106.900\pm0.11$ & $7.40$ & Background \\[2pt]
38 & HD224392 & 2 & 2005-11-07 & Ks & S27 & C & $4.0\times8\times10$ & $12.437\pm0.0237$ & $108.150\pm0.13$ & $10.8$ & Background \\
38 & HD224392 & 2 & 2008-08-21 & Ks & S27 & C & $4.0\times8\times10$ & $12.244\pm0.0159$ & $107.680\pm0.11$ & $10.7$ & Background \\[4pt]
\hline
\caption{\label{tab:cc} Multi-epoch measurement results for 41 companion candidates detected around 23 stars of the sample. The imaging mode used for each detection is indicated: (D) direct imaging, (S) saturated imaging (only on PUEO), (C) coronagraphic imaging (only on NaCO). The companion candidate status are defined in Sect.~\ref{sec:companionship}.\newline Notes for some targets: \newline HD~32743 --- $\rho$, $\theta$, and $\Delta m$ of CC\#2, 3, \& 4 have been determined with respect to the properties of CC\#1, which is seen both in direct and coronagraphic images.\newline HD~153363 --- Only the two brightest companion candidates are reported. Many other point sources ($\sim 56$) are seen using the coronagraph; however, they are all likely background contaminants as the star is close to the Galactic plane ($b=+10.6^\circ$). Besides, we miss a second epoch observation with the coronagraphic mask. \newline HD~216385 --- CC\#1 is not seen in the second epoch observation, while CC\#2 is too weak to have been detected during the first PUEO epoch. It was, however, out of the field of a third NaCo epoch not reported in this Table, and is thus a likely background contaminant.}
\end{longtable}
\end{landscape}}

\clearpage

\end{document}